\documentclass[%
 aip,
 amsmath,amssymb,
reprint,%
twocolumn
floatfix,
]{revtex4-1}
\usepackage{amsmath}
\usepackage[english]{babel}
\usepackage{appendix}
\usepackage{float}
\usepackage{graphicx}% Include figure files
\graphicspath{{./figures/}} % Trailing "/" is mandatory!!!

\usepackage[separate-uncertainty = true]{siunitx}
\usepackage{subcaption}
\usepackage{afterpage}
\usepackage{dcolumn}% Align table columns on decimal point
\usepackage{bm}% bold math
\usepackage{hyperref}% add hypertext capabilities
\DeclareSIUnit\dBm{dBm}
\DeclareSIUnit\dBc{dBc}
\newcommand{\chem}[1]{\ensuremath{\mathrm{#1}}}
\newcommand{\wn}[1]{\SI{#1}{\per\centi\meter}}
\newcommand{\ho}{\chem{Ho^{3+}}}
\newcommand{\lihof}{$\mathrm{LiYF}_4$:$\mathrm{\ho}$}
\newcommand{\etal}{\emph{et al.}~}
\newcommand{\thz}{\si{\tera\hertz}}

\newcommand{\affUCLEE}{Department of Electronic and Electrical Engineering, University College London,Torrington Place, London WC1E 7JE, United Kingdom}

\newcommand{\affLCN}{London Centre for Nanotechnology, University College London, 17-19 Gordon Street, London WC1H 0AH, UK}

\newcommand{\affETH}{Departments of Physics, ETH Z\"{u}rich, CH-8093 Z\"{u}rich, Switzerland and \'{E}cole Polytechnique F\'{e}d\'{e}rale de Lausanne (EPFL), CH-1015 Lausanne, Switzerland}
\newcommand{\affPSI}{Photon Science Division, Paul Scherrer Institute, CH-5232, Villigen, Switzerland}

\begin{document}

%\preprint{APS/123-QED}

\title{Optical communications-based platform uses Uni-Travelling-Carrier Photodiode for ultra-high resolution software-defined THz spectroscopy and reveals LiYF4:Ho intrinsic spectral line shape}
% Force line breaks with \\


\author{Rodolfo I. Hermans}
\email[]{r.hermans@ucl.ac.uk}
\affiliation{\affLCN}
\affiliation{\affUCLEE}

\author{Haymen Shams}
%\email[]{h.shams@ucl.ac.uk}
\affiliation{\affUCLEE}

\author{James P. Seddon}
%\email[]{james.seddon@ucl.ac.uk }
\affiliation{\affUCLEE}

\author{Alwyn J. Seeds}
%\email[]{a.seeds@ucl.ac.uk}
\affiliation{\affUCLEE}
\affiliation{\affLCN}

\author{Gabriel Aeppli}
%\email[]{gabriel.aeppli@psi.ch}
\affiliation{\affETH}
\affiliation{\affPSI}

%collaboration{CLEO Collaboration}%\noaffiliation

\date{\today}% It is always \today, today, but any date may be explicitly specified

\begin{abstract}
High resolution (900 Hz full-width at half maximum) frequency domain spectroscopy near \SI{0.2}{\tera\hertz} is achieved using an exact frequency spacing comb-source in the optical communications band, filtering and photo-mixing in a custom Uni-Travelling-Carrier Photodiode (UTC-PD) for THz signal generation and coherent down-conversion for detection. Via time domain modulation of one of the comb lines, a fully controllable spectrometer can be defined in software, and this principle is demonstrated for magnetic field-free readout of the electronuclear spectrum for the Ho ions in \lihof, a material often used for demonstration experiments in quantum science. In particular, homogeneous and inhomogeneous contributions to the spectrum are readily separated.

\end{abstract}

\maketitle
%\tableofcontents

% Sub-Sections and lower hierarchy titles are for structure only and may not appear in actual paper.

\section{\label{sec:intro}Introduction}

%%% Figure Experiment schematics
\begin{figure*}[tbh]
\centering
\includegraphics[width=\textwidth]{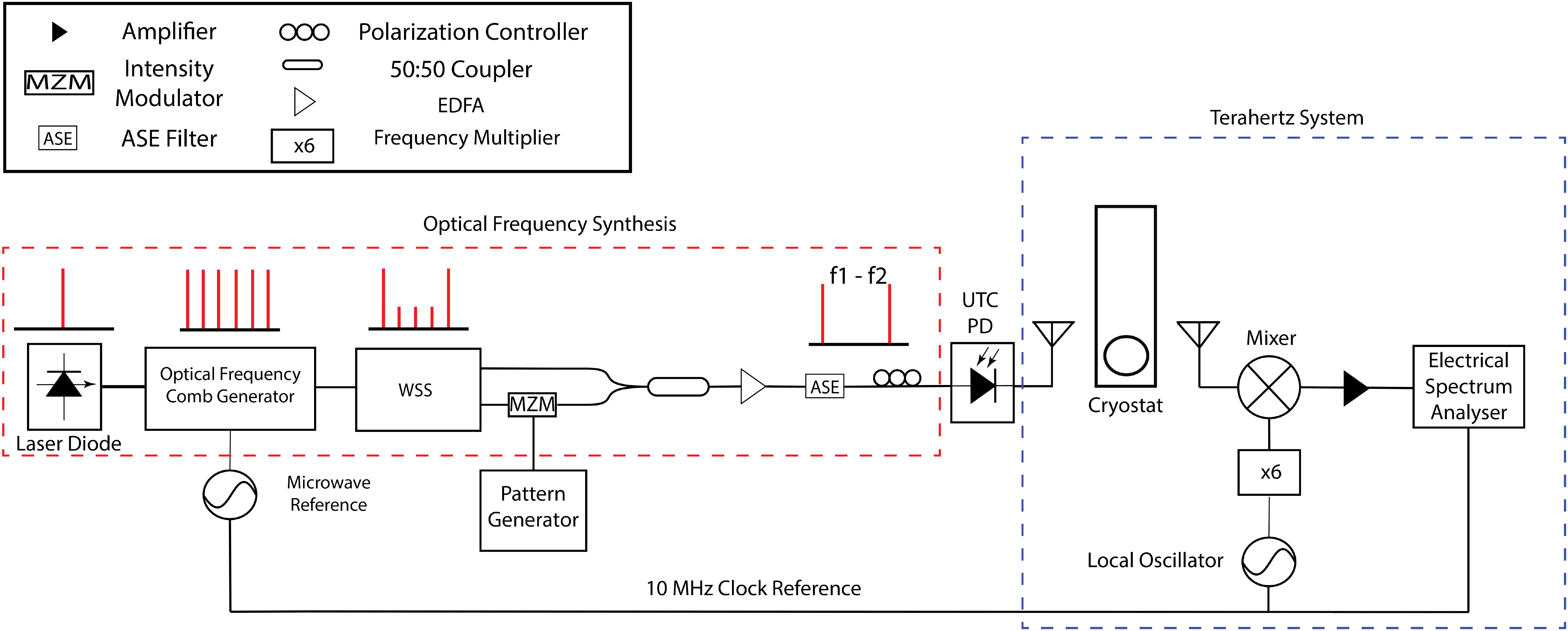}
\caption{Experiment schematics: A monochromatic telecommunications wavelength laser, feeds an optical frequency comb generator (OFCG) with exact tuneable spacing. A programmable wavelength selective switch (WSS) selects two bands 12 peaks apart, amplified and mixed in uni-travelling-carrier photo-diode (UTC-PD). A \SI{200}{\giga\hertz} beat frequency is transmitted through horn antennas and lenses though a continuous-flow liquid helium cryostat with thin polypropylene windows and \lihof sample. The received signal is down-converted using a sub-harmonic mixer and measured using a microwave spectrum analyser.}
\label{fig:ExperimentSchem}
\end{figure*}

The Terahertz (\thz) electromagnetic band (\SIrange{0.1}{10}{\tera\hertz}) lies in the technically challenging spectral gap between infrared light and microwave radiation\cite{Dhillon_2017,Seeds2013}. Specialised technology for the generation, manipulation and detection of so-called `T-rays', has come later than for other frequency bands and is still an active area of development.

The THz region offers many opportunities. For example, THz radiation is mainly innocuous to live tissue~\cite{Hintzsche2012}, making it an attractive option for medical diagnostics and material identification for safety and  security.
Despite the significant absorption of THz radiation by the atmosphere, the existing transmission windows also allow useful applications both in astronomy~\cite{Graf2015} and telecommunications\cite{Seeds2015}. Finally, various THz excitations in solids offer promise for quantum technology\cite{Greenland_2010}, if only compact coherent sources were available.

THz spectroscopy is currently dominated by Fourier Transform Infrared (FTIR)\cite{IRSpecRev} and Time Domain Spectroscopy (TDS). Off-the-shelf FTIR such as \emph{Bruker IFS 125HR} FTIR Spectrometer claims broad spectral range from \wn{5} (\SI{2}{\milli\meter}, \SI{149.9}{\giga\hertz}) in the far-IR to \wn{50000} (\SI{200}{\nano\meter}; \SI{1.499}{\peta\hertz}) in the UV and very high resolution down to \wn{9E-4} (\SI{26.98}{\mega\hertz})\cite{BrukerWebIF125SHR}.
TDS can achieve \SI{3}{\tera\hertz} bandwidth and \SI{1}{\giga\hertz} resolution via  Asynchronous Optical Sampling (ASOPS)~\cite{Elzinga1987,Bartels2007}.

However, THz Frequency domain spectroscopy is comparably less technologically mature, with few commercial systems produced~\cite{Stanze2011}. Typically, these systems make use of free running lasers, where the optically generated THz linewidth, and hence spectral resolution, is limited by the convolution of the two free running laser tones\cite{8388199}.

A spectroscopically measured absorbance curve depends on the convolution of both the sample absorption and the spectrometer response.
De-convolution is possible, but in most cases it is desirable to resolve the sample response without relying on noise-limited deconvolution processes\cite{Kauppinen1981}.
Resolving a narrow spectral feature normally requires even sharper discrimination either in the instrumentation illumination or detection.

In this work we exploit a method for narrow linewidth generation of coherent THz photons to introduce a new platform for spectroscopy. By controlling the illumination spectral line shape (SLS), we can gather information on the sample intrinsic SLSs. We demonstrate this technique by providing high resolution spectroscopy transmission measurements on \lihof, a material of interest for quantum science and technology, where we discriminate between inhomogeneous Gaussian and homogenous Lorentzian contributions to the absorption lines near \SI{0.2}{\tera\hertz}.

\subsection{Coherent Photonic THz Generation}
A particularly promising means of producing spectrally pure continuous wave (CW) THz signals is through the use of  the optical telecommunications toolbox~\cite{Nagatsuma2016}, containing a large variety of readily available and low cost photonic components, and ultra-fast Indium Phosphide (InP) photodiodes.

%\paragraph{UTC-PD as THz source}

The introduction of ultra-fast photodiodes such as Uni-Traveling-Carrier Photodiodes (UTC-PD)\cite{FirstUTCPD, Rouvalis2011}, initially for high-speed telecommunications, has allowed the generation of millimetre waves through photo-mixing. UTC-PDs mitigate the transit time response limited bandwidth of pin photodiodes through blocking hole injection and facilitating the injection of hot electrons to the sweep-out layer. UTC-PD frequency response is only limited by the electron transit and RC time constants of the photodiode.

Early UTC-PD devices were vertically illuminated structures with a trade-off between optical responsivity and bandwidth~\cite{LampinCavityUTC}. In this geometry the transit time response is improved by reduction of the absorber layer thickness. This reduction comes at the cost of reduced optical absorption in the photodiode absorption layer. There have been several structures that attempt to resolve this through the introduction of optical cavities~\cite{LampinCavityUTC} at the expense of more challenging fabrication.

Alternatively, the bandwidth responsivity trade-off can be mitigated through the integration of the UTC-PD structure within an optical waveguide coupled photodiode~\cite{renaud2006high}. In this case the photodiode is evanescently coupled to a passive optical waveguide where the optical absorption and carrier transit direction are perpendicular to each other. This enables a thinner absorber layer to be used without a reduction in optical responsivity. Waveguide to chip coupling can be improved using optical mode converters to improve further the optical responsivity of such edge coupled devices~\cite{Rouvalis2010}.

%\paragraph{Free lasers vs comb source}

Optical heterodyne generation of THz involves the beating of two optical tones $f_1$ and $f_2$ in a high speed photodiode such as an UTC-PD. % discussed previously.
The resulting THz signal includes a strong component at the heterodyne frequency $|f_2-f_1|$ with a spectral line-shape given by  the convolution of the two free running laser emission spectra functions. The THz signal can be tuned by varying $f_1$, $f_2$ or both. %
Typical semiconductor lasers have a significant linewidth due to fluctuations of the effective cavity length. Linewidths vary from \SI{10}{\kilo\hertz} in the case of Distributed Feed Back (DFB) Lasers to \SI{100}{\kilo\hertz} for widely tuneable external cavity lasers\cite{8388199}.

To achieve the spectrally pure signals needed to resolve details such as the SLS of hyperfine structure and gas absorption lines we can exploit an optically phase correlated source for optical heterodyne generation of THz signals.
This can be achieved through the use of optical frequency comb which generates a series of phase correlated lines with discrete frequency spacing.

Several methods have been proposed for the generation of Optical Frequency Combs (OFCs), for example: Kerr combs using micro-resonators\cite{del2007optical,Kues2019}, mode-locked lasers\cite{MLLComb}, and fibre optic modulators.
OFCs employing Lithium Niobate modulators~\cite{ModulatorComb, Bennett1999, monolithComb} are attractive as comb sources due to the tuneability of the centre wavelength, a greater accuracy through referencing of the comb generator to a supplied reference frequency and increased frequency precision.
Use of a modulator in a recirculating fibre loop with erbium doped fibre amplification (EDFA) allows for comb spans of up to greater than \SI{2.7}{\tera\hertz}~\cite{PonnampalamComb}. Narrow span combs with maximum flatness have also been achieved initially through cascaded Mach Zehnder and phase modulators~\cite{monolithComb, Takita:04}.

Comb generation can be achieved in a more compact form by the use of a dual drive Mach Zehnder modulator, with comb spans up to \SI{0.23}{\tera\hertz} being demonstrated~\cite{SakamotoComb}.
In this case the construction of the comb is compact and comprised of off-the-shelf telecommunications wavelength components and achieves a comb span suitable for accessing our region and resolution of interest i.e. the lower lying spectral features of \texorpdfstring{\lihof}{LiYF4:Ho} around \SI{0.2}{\tera\hertz}.

\subsection{Sample}
We report the first high resolution direct optical measurements of the lowest THz transition on \ho~ions in \lihof. Single crystal Yttrium Lithium Fluoride (YLF, \chem{LiYF_4}) features a scheelite-structure,
tetragonal space group $\mathrm{C_{4h}^6}$. Rare earth dopant substituting \chem{Y^{3+}} at sites with $\mathrm{S_4}$ symmetry cause insignificant crystalline structure perturbation~\cite{PhysRevB.13.2805} and have attracted considerable enthusiasm from both the optics and quantum science and technology communities.  From the potential practical use in laser gain mediums~\cite{Santo2006b},
to a platform for fundamental research on Ising ferromagnetism~\cite{Reich1990},
quantum annealing~\cite{Brooke1999,Ronnow2007},
quantum tunnelling of magnetic domain walls~\cite{Brooke2001},
coherent spin oscillations (Rabi)~\cite{Ghosh2002},
entangled quantum states of magnetic dipoles~\cite{Ghosh2003a},
and quantum phase transitions~\cite{Bitko1996,Ronnow2005}.
Rare earth doped materials have also been identified as candidates for quantum information applications~\cite{Bertaina2007, Bussieres2014}.
There is particular interest in systems where electronuclear qubits can be read using optical transitions\cite{Rancic2017}.
The $4f^{10}$ electronic configuration of \ho~is split in three by the Coulomb field and in five by spin-orbit coupling. We focus on the lower spin-orbit state $^5I_8$ which is further split by the crystal field. Research by Karayianis identified this lowest levels as $\Gamma_{3,4}$ doublet and two first excited states as $\Gamma_{2}$ singlets with \SI{0.2}{\tera\hertz} (\wn{7}) and \SI{0.7}{\tera\hertz} (\wn{23})~\cite{Karayianis1976}. Electron-paramagnetic-resonance experiments (EPR) by \mbox{Magari{\~n}o} determined the matrix elements coupling them~\cite{PhysRevB.13.2805,PhysRevB.21.18}. We measure the \SI{0.2}{\tera\hertz} transition from the $\Gamma_{3,4}$ doublet ground state to the first excited $\Gamma_{2}$ state.

%\begin{table}
%\begin{tabular}{|c|c|c|c|c|}
%  \hline
%  ~ & Concentration [\%]& L1 [mm] & L2  [mm]& L3  [mm]\\   \hline
%  Sample 1 & 0.10 & 5 & 5 & 19.8 \\
%  Sample 2 & 0.25 & 5 & 5 & 20.0 \\
%  Sample 3 & 0.50 & 5 & 5 & 20.0 \\
%  Sample 4 & 1.00 & 10 & 10 & 2.18 \\
%  Sample 5 & 8.0 & ? & ? & ? \\
%  \hline
%\end{tabular}
%\caption{Geometry and \ho concentration of samples}
%\label{table:samples}
%\end{table}

\section{\label{sec:ProgSpec}Programmable spectroscopy}

We introduce the concept of \emph{software-defined spectroscopy} as the exploitation of fast modulation of optical or RF signals in order to create illumination with bespoke spectral curves optimised to extract specific features of sample optical response. Specific modulation patterns applied to simpler signals  exploits common telecommunications technology to allow new and inexpensive approaches to access physical parameters that could circumvent common experimental limitations. Formally, as shown in Fig. 1, the
mixing diode multiplies two optical communications inputs, one modulated by the pattern generator, to produce THz radiation

\begin{figure}[ht!]
\includegraphics[width=0.9\columnwidth]{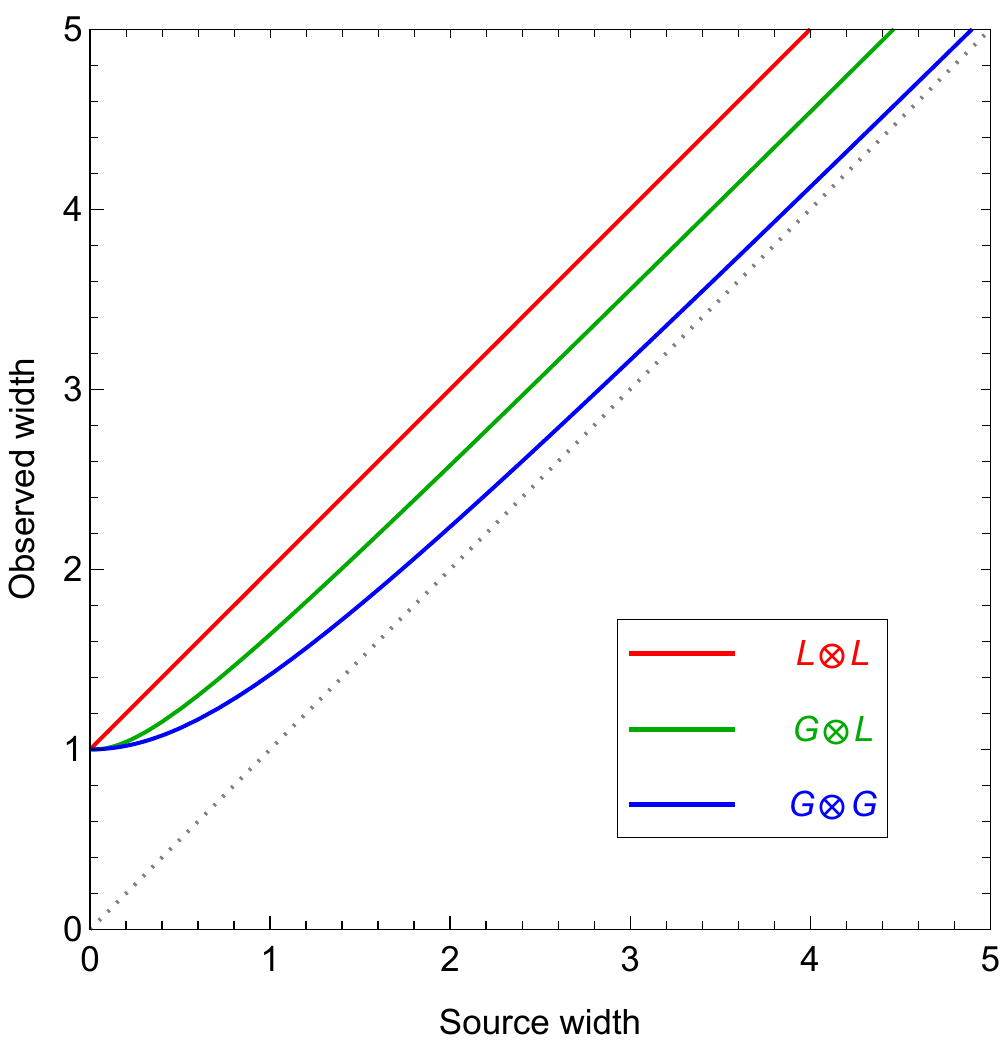}
\caption{Expected FWHM for three different convolution tuples. The convolution of two Lorentzians curves is also a Lorentzian with FWHM equal to the sum of its components and therefore appears as linear curve (red). The convolution of two Gaussians is also a Gaussian curve with FWHM given by the RMS (blue). The convolution of Gaussian and Lorentzian curves gives a Voigt profile with FWHM in between (green).  }
\label{fig:voigt}
\end{figure}

\subsection{Implementation of digital modulation}
We use a particularly simple example of software-defined spectral modification with a random on-off keying (OOK) modulation.
The generation of the specific modulation patterns was achieved through the use of an inexpensive fibre optic intensity modulator (bandwidth < \SI{10}{\giga\hertz}) placed on one of the filtered comb lines. By applying a pseudo-random bit sequence generated from a pulse pattern generator, arbitrary broadening of the laser tone is achieved.
The resulting spectra can be approximated as a Gaussian broadening of the spectrum with a width defined by the time-base of the pulse pattern.

\subsection{Possible spectral shapes}
A more generalised broadening can be achieved when the OOK is defined as a two state discrete Markov stochastic process defined by the transition matrix $M$ and probability $p$.

\begin{equation}\label{eq:stocmatrix}
M = \left(
\begin{array}{cc}
 p & 1-p \\
 1-p & p \\
\end{array}
\right)
\end{equation}

We derive in the supplementary material the exact analytical form of the spectra for a Markov OOK by calculating the correlation function and using the Wiener-Khinchin-Einstein theorem~\cite{Khintchine1934}. We recognize three cases that are of particular interest. Consider first $0<p<1/2$ with delaying correlation $\exp{(\log(1-2p) t/\tau)}$ and an exact Lorentzian Power Spectral Density function  (PSD). Second, an unbiased pseudo-random bit sequence corresponding to a Markov process with $p=1/2$ well approximated by a Gaussian PSD, with $\sigma = \sqrt{6}/\tau$. Finally, the case $1/2< p <1$ with oscillatory correlation $\exp{(\log(1-2p) t/\tau)} \cos (\pi t /\tau)$ featuring a side peak.
By this technique, the modulated signal will be broadened by a convoluting spectrum that depends not only on the pattern time-base but also in digitally defined OOK defined by a state transition probability.

The digital OOK pattern is not limited to a Markovian process, but could be any arbitrary bit sequence pattern. Here we focus only on exploiting the ability to generate both Gaussian and Lorentzian SLS.

It follows from the convolution theorem that the spectrum of the modulated signal is the convolution of the original signal with the spectrum of the modulation. Consequently, the illumination SLS can always be broadened with a selection of shapes defined by the stochastic OOK. Both Gaussian and Lorentzian lineshapes can be easily obtained by tuning the value of $p$ defining the Markovian  OOK.

\subsection{Application}

We are interested in  obtaining the intrinsic SLS of the sample distinguishing the Lorentzian homogeneous contribution from the Gaussian inhomogeneous contribution~\cite{koechner2013solid}. We first consider the convolution of tuples of Gaussians $G(\mu,\sigma)$ and Lorentzians  $L(\mu, \gamma)$.

\begin{subequations}
\begin{eqnarray}
  G(\mu_1,\sigma_1)  \otimes G(\mu_2,\sigma_2)  &=& G(\mu_1+\mu_2, \sqrt{{\sigma_{1}}^2+{\sigma_{2}}^2})\\
  L(\mu_1, \gamma_1) \otimes L(\mu_2, \gamma_2) &=& L(\mu_1+\mu_2, \gamma_1+\gamma_2)\\
  L(\mu_1, \gamma)   \otimes G(\mu_2,\sigma)    &=& V(\mu_1+\mu_2, \gamma,\sigma)
\label{eq:voigt}
\end{eqnarray}
\end{subequations}

%where the Gaussian $G(x,\mu,\sigma) = \frac{1}{\sqrt{2 \pi } \sigma } e^{-\frac{(x-\mu )^2}{2 \sigma ^2}}$ , the Lorentzian $L(x, x_0, \gamma) = \frac{\gamma}{2\pi  \left(\frac{\gamma^2}{4}+{\left(x-x_0\right)}^2\right)}$ and $V(x,\mu_1+\mu_2, \gamma,\sigma)$ is the Voigt profile.

where $V(\mu_1+\mu_2, \gamma,\sigma)$ is the Voigt profile. The corresponding full width at half maximum (FWHM) are

\begin{subequations}
\begin{eqnarray}
  \Gamma_{\text{G,G}} &=& 2\, \sqrt{{\sigma_{1}}^2+{\sigma_{2}}^2} \sqrt{2 \ln{2}}  \label{eq:FWHMgg} \\
  \Gamma_{\text{L,L}} &=& 2\, (\gamma_1+\gamma_2)  \label{eq:FWHMll} \\
  \Gamma_{\text{L,G}} &\approx& \sqrt{0.8664 \gamma ^2+8 \sigma ^2 \log (2)}+1.0692  \label{eq:FWHMlg} \gamma
\end{eqnarray}
\end{subequations}

where the last equation is an approximation~\cite{OLIVERO1977233}. Equations \ref{eq:FWHMgg}-\ref{eq:FWHMlg} and Figure \ref{fig:voigt} reveal three distinguishable FWHM trends: a linear behaviour for Lorentzian, Root Mean Square (RMS) for Gaussian and a more complex expression in between the former two for Voigt~\cite{HE20135245}. This observation opens the possibilities of identifying a SLS that is not resolvable by the spectrometer resolution by examining the FWHM trend by well controlled illumination SLSs.

\section{\label{sec:methods}Methods}

\begin{figure}[ht!]
\begin{subfigure}{\columnwidth}
\includegraphics[width=0.95\columnwidth]{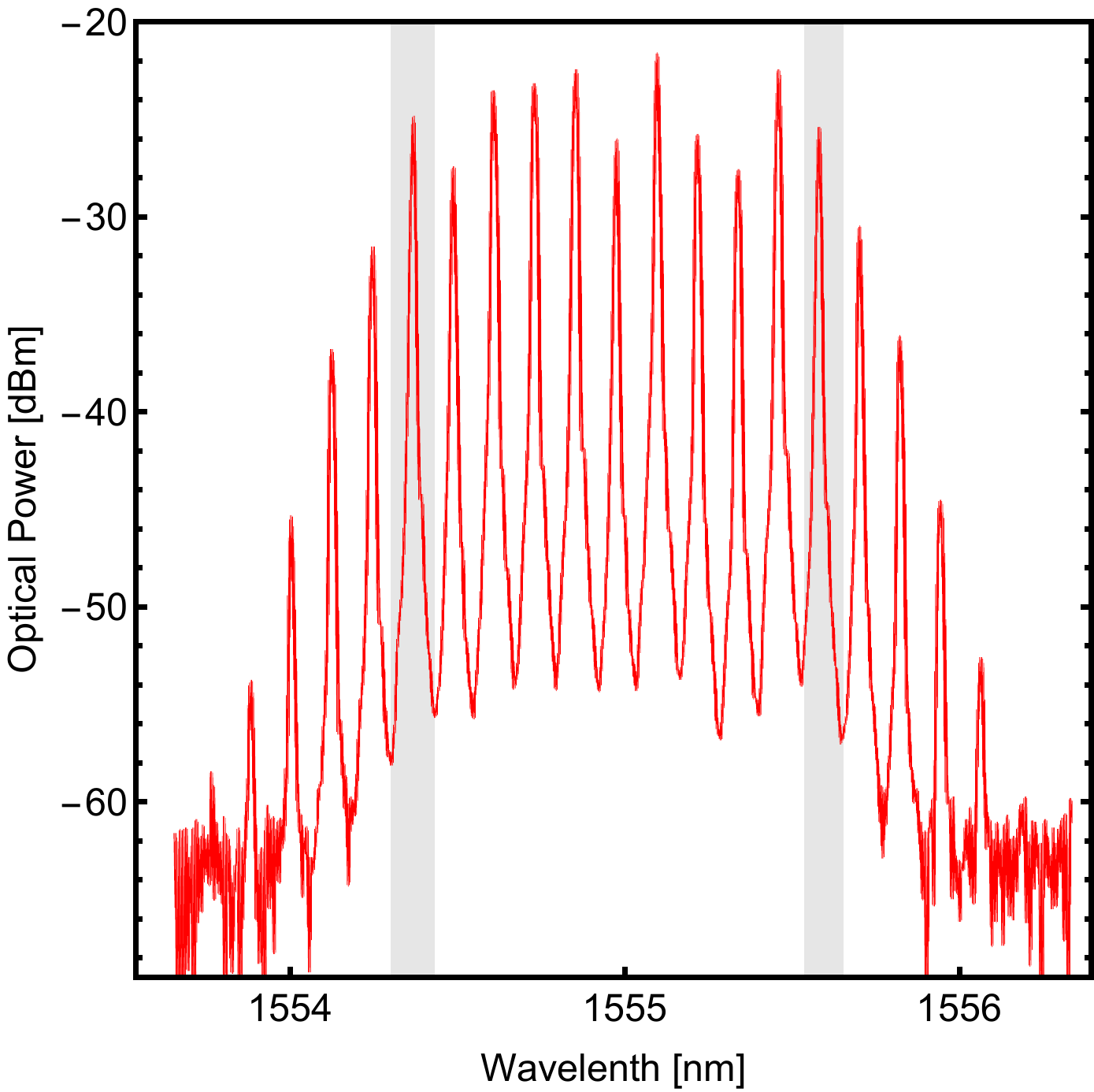}
\caption{Optical spectrum of the synthesised comb with tuneable spacing (here \SI{15}{\giga\hertz}), the highlighted peaks are selected with a Waveshaper and then amplified, filtered and mixed in a UTC-PD.}
\label{fig:CombSpectrum}
\end{subfigure}
\begin{subfigure}{\columnwidth}
\includegraphics[width=0.95\columnwidth]{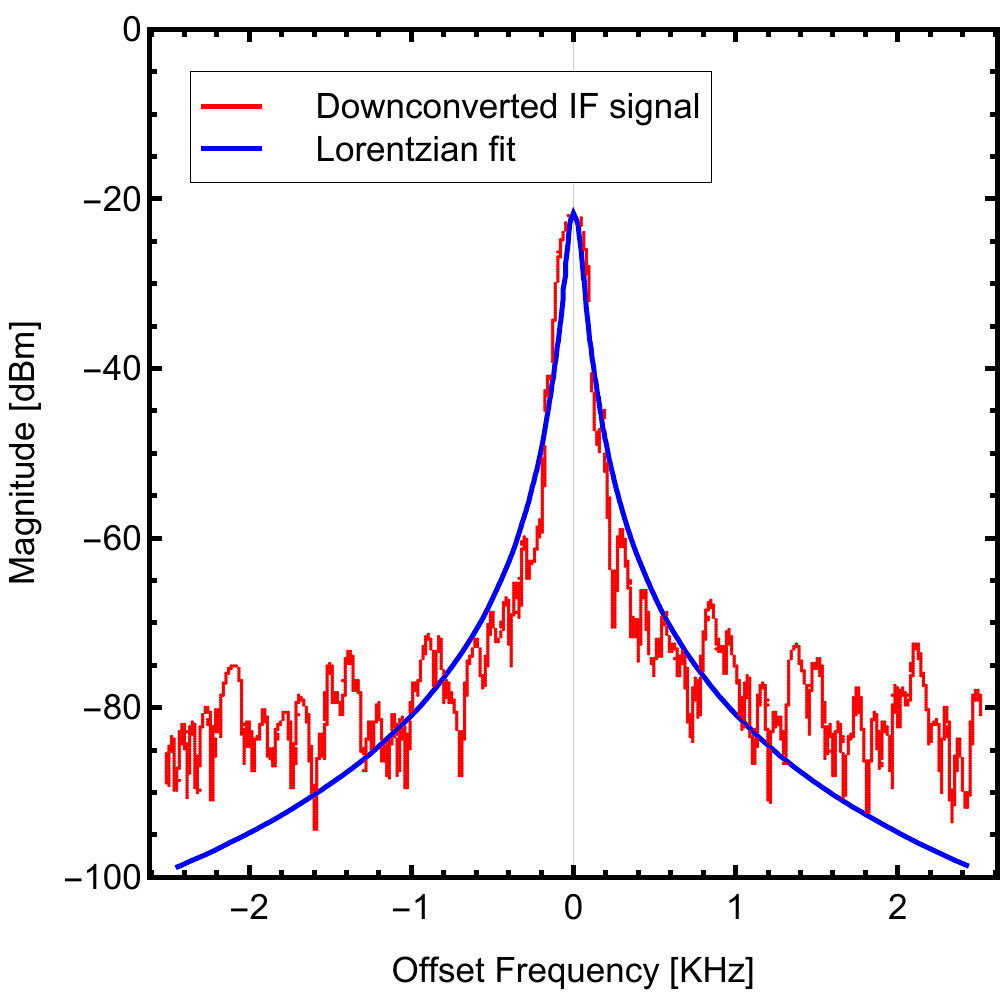}
\caption{Down converted IF signal from \SI{246}{\giga\hertz} as measured by Agilent 8565E spectrum analyser. The peak value is stored as the transmitted amplitude. Centre frequency \SI{15}{\giga\hertz}}
\label{fig:IFTrace}
\end{subfigure}
\caption{Signal generation and detection. }
\end{figure}

The experimental arrangement is shown in Figure~\ref{fig:ExperimentSchem}.

\subsection{OFC}

Frequency synthesis was carried out using an OFC generator detailed by Shams~\etal~\cite{Shams:16}. The comb is formed via modulation of a single \SI{10}{\kilo\hertz} linewidth laser signal at \SI{1553.7}{\nano\meter} (RIO Orion Series) by a microwave synthesiser \emph{Rohde\&Schwarz SMP04 \SI{40}{\giga\hertz}}. The modulator used is a dual drive Mach Zehnder, which generates an optical comb spectrum with a flat response~\cite{SakamotoComb}. Comb line selection is achieved using a multi-port programmable optical filter (Finisar Waveshaper 4000S) insertion loss \SI{4.5\pm0.1}{\dB} resulting in two coherent lines with \SI{30}{\dB}  suppression of adjacent comb lines.

The two selected comb lines are amplified using an Erbium Doped Fibre Amplifier (EDFA), with a band pass filter placed after the EDFA to reduce noise contributions from Amplified Spontaneous Emission (ASE). To achieve high resolution tuning, the microwave synthesizer frequency is tuned across the frequency span between comb lines followed by sequentially shifting the programmable optical filtering band.
A pause is introduced after changing filter settings with a settling time of \SI{500}{\milli\second}. Figure \ref{fig:CombSpectrum} shows the synthesised OFC spectrum with a comb span of \SI{0.27}{\tera\hertz}, a comb line spacing of \SI{15}{\giga\hertz} and two selected comb lines for \SI{150}{\giga\hertz} beat frequency.

\subsection{UTC}

The UTC-PDs used in this study were grown by gas source molecular beam epitaxy by III-V lab, the epitaxial structure of the devices is detailed in~\cite{Rouvalis:12}. The device geometry was a $2 \times \SI{15}{\micro\meter\squared}$ with a \SI{3}{\dB} bandwidth of \SI{90}{\giga\hertz}, and optical responsivity of 0.2 A/W. The UTC-PD was integrated with a co-planar waveguide, the THz signal was extracted using a \emph{Cascade Microtech Air Co-Planar} (ACP) probe with a bandwidth of \SIrange{140}{220}{\giga\hertz}. Flann 20 dB standard gain horn antennas (\SIrange{145}{220}{\giga\hertz} bandwidth)  were used to form the free space path within which the sample was placed.

\subsection{Detection}
The THz signal is detected using a sub harmonic mixer (VDI WR5.1 SHM) with a bandwidth of \SIrange{140}{220}{\giga\hertz}, driven by a $\times6$ frequency multiplier (OML S10MS-AG)  supplying a Local Oscillator (LO) signal between \SIrange{70}{110}{\giga\hertz}. The multiplied LO signal is supplied from an Agilent E8257D \SI{65}{\giga\hertz} signal generator. The THz signal was down-converted to a \SI{15}{\giga\hertz} Intermediate Frequency (IF) (See figure~\ref{fig:IFTrace}) and recorded on an Agilent 8565E spectrum analyser. The down-converted IF signal linewidth was \SI{900}{\hertz}, measured using the electrical spectrum analyser.  A common \SI{10}{\mega\hertz} reference clock is supplied to the LO signal generator and spectrum analyser from the comb line signal generator.

\subsection{Sample}

The sample of \lihof was commercially provided by UAB ALTECHNA at 1\% \ho concentration custom dimensions $\SI{10}{\milli\meter}\times \SI{10}{\milli\meter}\times \SI{2.18}{\milli\meter}$ crystals were cut and optically polished with c-axis in plane. The test sample was mounted on the cold finger of a continuous flow liquid helium cryostat such that THz radiation could not be transmitted around the sample. The THz transmission through the sample was linearly polarised with the electrical field along the \emph{c} crystalline axis.
The cooled sample was illuminated through thin polypropylene windows, liquid helium was pulled by a diaphragm vacuum pump and the sample temperature was maintained at \SI{4}{\kelvin} or \SI{200}{\kelvin} by an Oxford ITC4 controlling a resistive heater.

\section{\label{sec:results}Results}

\begin{figure}[h]
\begin{subfigure}{0.9\columnwidth}
\includegraphics[width=\columnwidth]{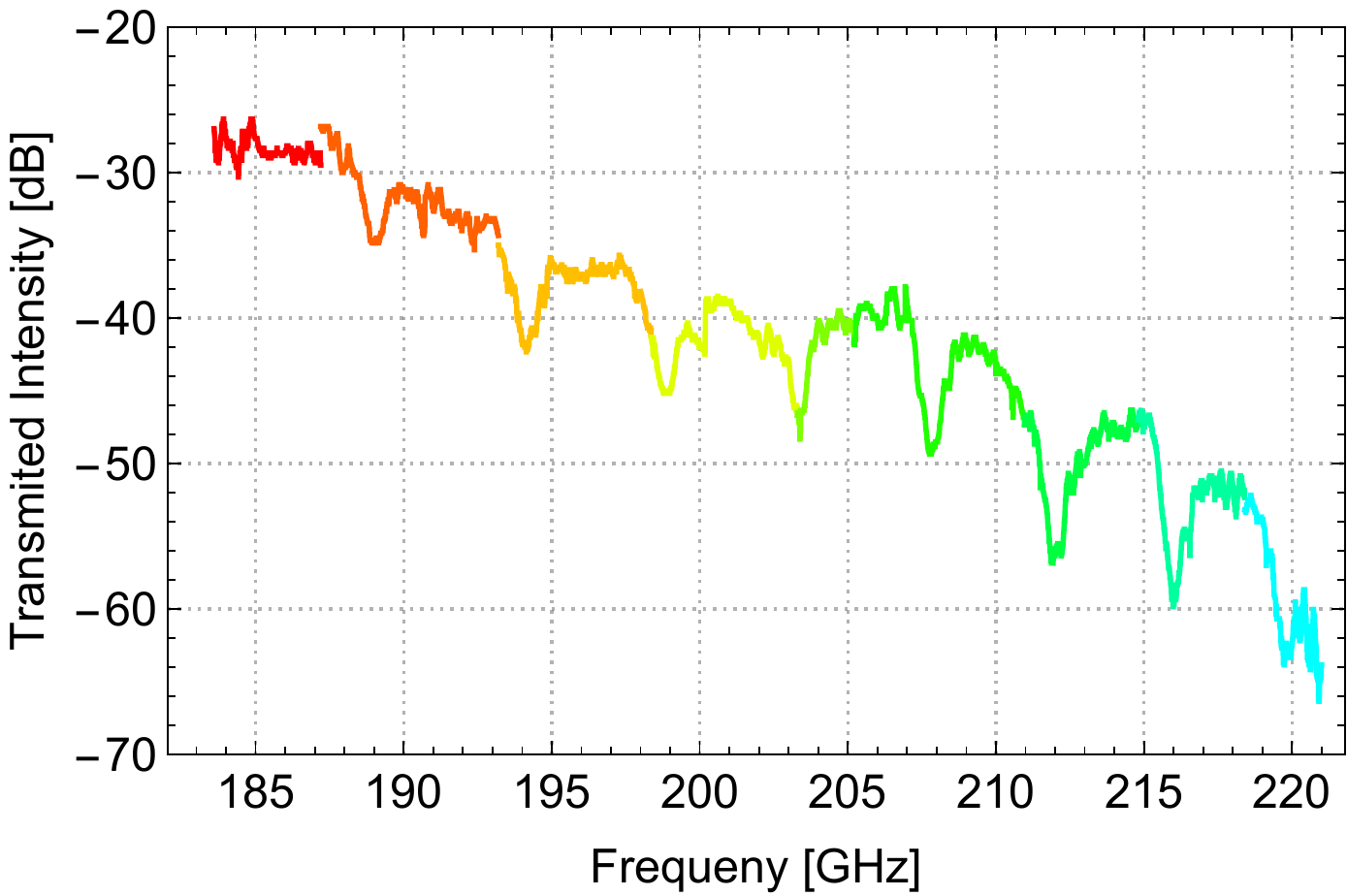}
\caption{Raw data composite.}
\label{fig:RawDataComposite}
\end{subfigure}
\begin{subfigure}{0.9\columnwidth}
\includegraphics[width=\columnwidth]{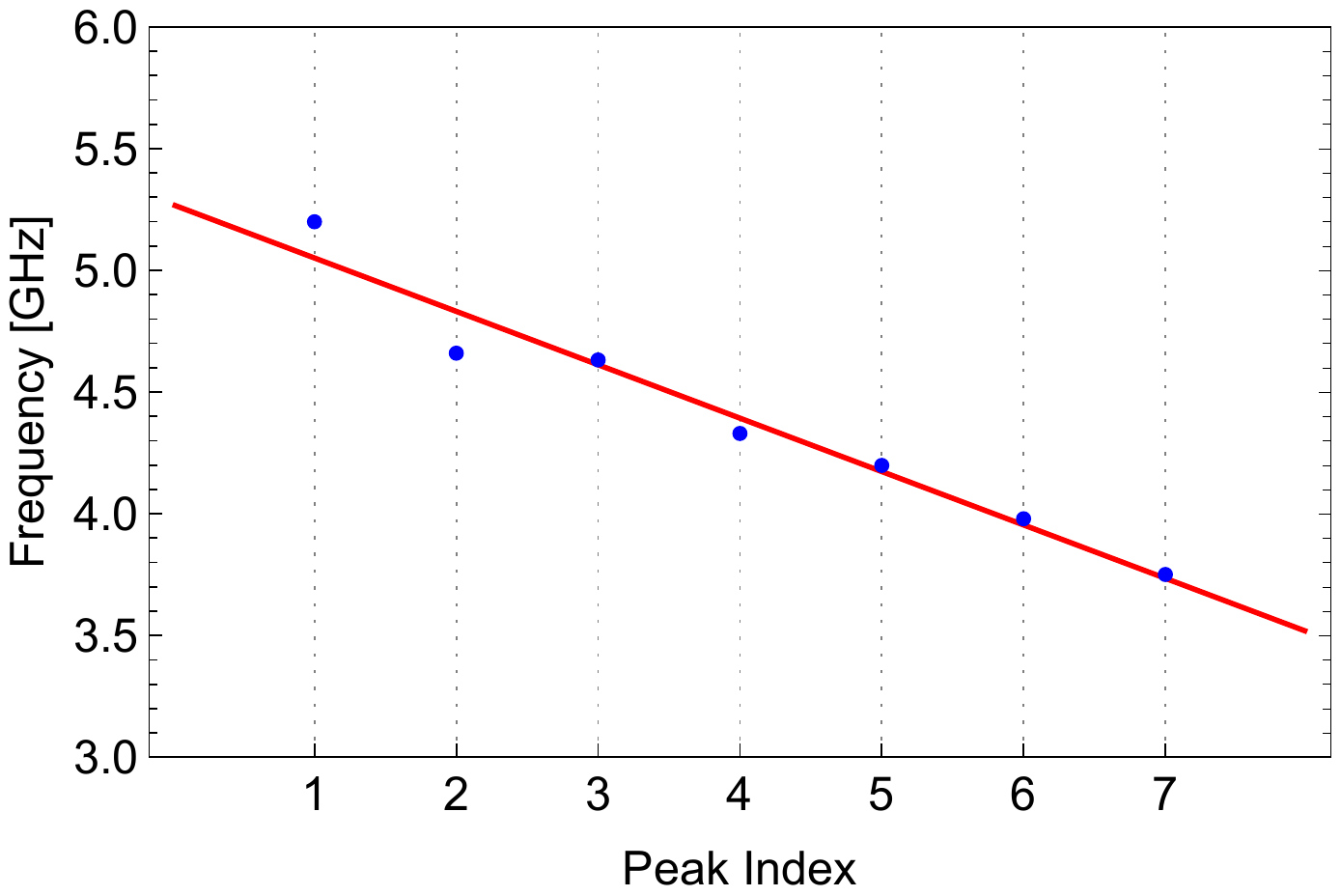}
\caption{Frequency differences. }
\label{fig:spacingtrend}
\end{subfigure}
\caption{\lihof absorption data. a) Raw data composite of measured transmitted intensity. Each colour corresponds to the set of data points collected with the same waveshaper configuration. The downwards trend corresponds to the emission and detection attenuation. The big dips correspond to the \lihof sample absorptions. b) Frequency difference between consecutive Hyperfine absorbance peaks is well described by a linear fit with slope \SI{-0.22\pm0.02}{\giga\hertz} }
\end{figure}

%\subsection{Raw spectrum}
The composite raw spectral data acquired for the \lihof at \SI{4}{\kelvin} is shown in Figure \ref{fig:RawDataComposite}. This corresponds to the measured transmitted power normalised by the measured photocurrent. Each colour indicates the range of data acquired for a single configuration of the waveshaper. The strong overall decaying trend is attributed to the UTC-PD performance and bandwidth limitations imposed by the ACP Probe. The dips are attributed to the absorption lines of the sample. A similar curve is later obtained for the same conditions except the \lihof sample is at \SI{200}{\kelvin} and used as a reference for normalisation and to estimate absorbance curves. At \SI{200}{\kelvin} absorption peaks are no longer measurable as higher states are thermally populated and the weak absorption peaks have been flattened by Doppler broadening.

\subsection{Observed absorbance}
We observe 8 absorption lines in the range between \SIrange{189}{220}{\giga\hertz} with linearly decreasing spacing ranging from \SIrange{5.2}{3.75}{\giga\hertz}. Table~\ref{tab:peaks} shows the peak values positions and differences and Figure~\ref{fig:spacingtrend} shows the spacing linear trend. The absorbance line shape appears broader than the experimental resolution so we seek confirmation for the line shape characteristics.

\begin{table}[bth]
\begin{tabular}{c c c c}
\,N\, & \,Frequency [\si{\giga\hertz}]\, & \,Absorbance\, & \,Difference  [\si{\giga\hertz}]\,\\
\hline
1 & 189.00 & 0.29 & 5.20 \\
2 & 194.20 & 0.37 & 4.66 \\
3 & 198.86 & 0.43 & 4.63 \\
4 & 203.49 & 0.37 & 4.33 \\
5 & 207.82 & 0.49 & 4.20 \\
6 & 212.02 & 0.39 & 3.98 \\
7 & 216.00 & 0.48 & 3.75 \\
8 & 219.75 & 0.35 & -    \\
\end{tabular}
\caption{Summary of absorption peak data}
\label{tab:peaks}
\end{table}

\subsection{Peak broadening}

\begin{figure}[th!]
\centering
\includegraphics[width=0.9\columnwidth]{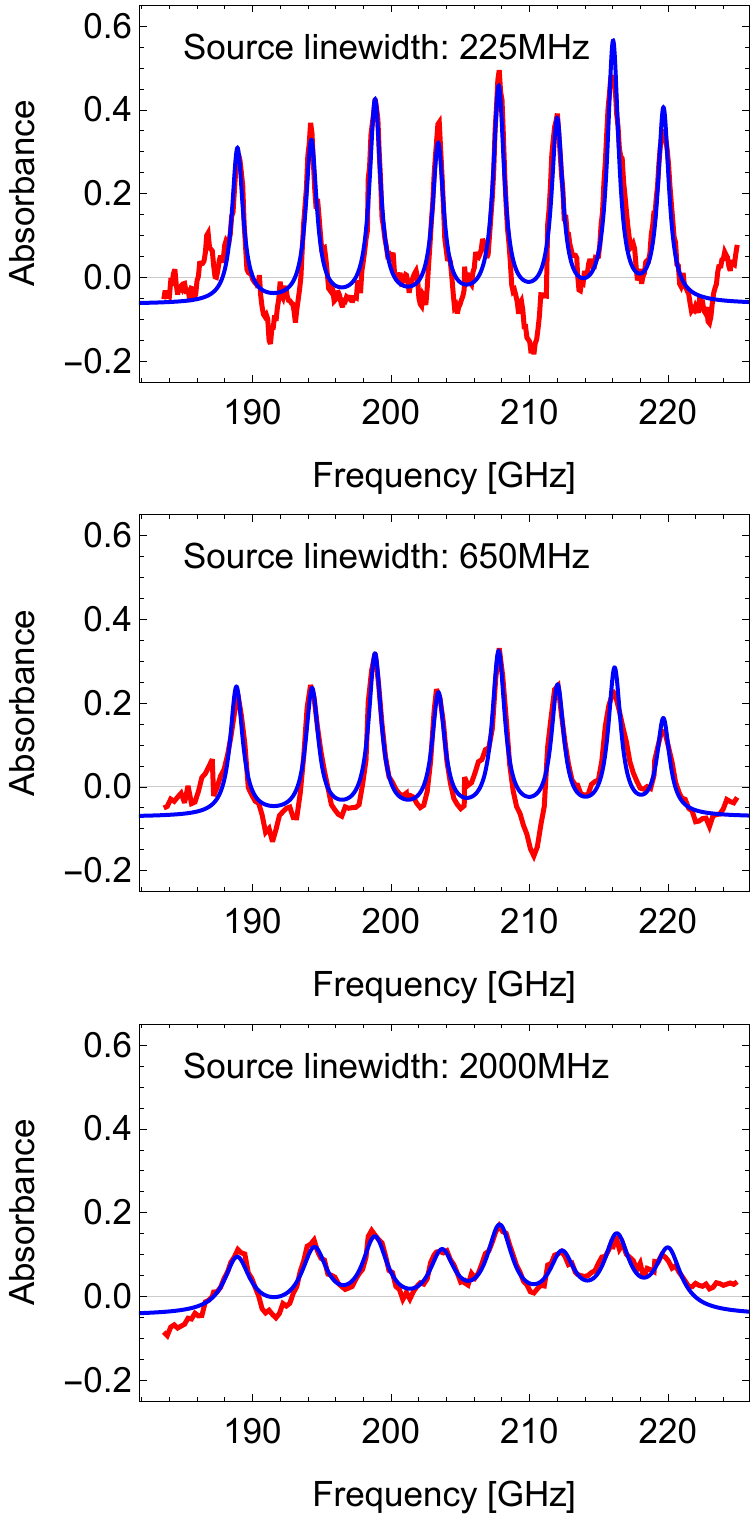}
\caption{Gallery of absorbance curves and respective fitting functions for a selections of illuminations with different spectral line-widths. As the illumination spectra broadens so does the estimates absorbance spectra line shape. The shape of the observed absorbance line is the convolution of the illumination spectrum and the intrinsic absorption curve.}
\label{fig:gallery}
\end{figure}

We use a Gaussian broadening of the illuminating spectral shape line and measure the FWHM of the estimated absorbance curves as a function of the illumination linewidth. Figure~\ref{fig:gallery} show three representative cases for source linewidth of \SI{225}{\mega\hertz}, \SI{650}{\mega\hertz} and \SI{2000}{\mega\hertz} and the respective Gaussian fits.

\subsection{Convolutions tuples}

\begin{figure}[th!]
\centering
\includegraphics[width=0.9\columnwidth]{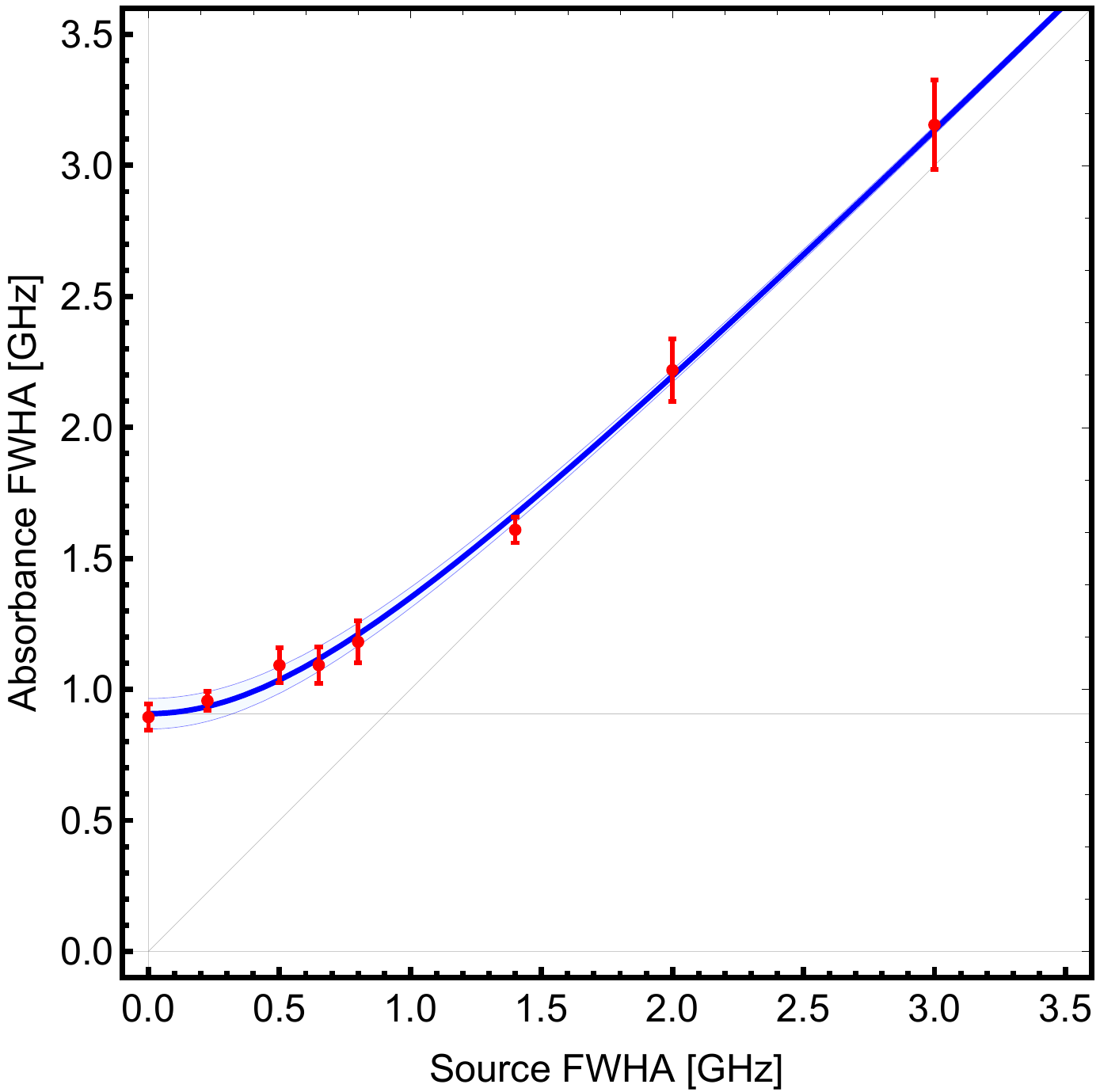}
\caption{Measured absorbance FWHM ($w_{m}$) illumination FWHM ($w_{i}$) is well fitted by an RMS model $\sqrt{w_{i}^{2}+w_{s}^{2}}$, with a sample FWHM $w_s=\SI{0.907\pm0.017}{\giga\hertz}$. The plot includes a very narrow \SI{99}{\percent} confidence band. We interpret this result as a clear signature that both source and sample intrinsic SLSs are Gaussian.}
\label{fig:broadening}
\end{figure}

The absorbance FWHM increases with the illumination broadening revealing an RMS trend as expected for the convolutions of two Gaussian SLSs.
Figure~\ref{fig:broadening} shows a good RMS fit ($w_m= \sqrt{w_{i}^{2}+w_{s}^{2}}$) of the absorbance FWHM as a function of illumination FWHM ($w_{i}$) revealing a good estimation for the sample linewidth  $w_{s}=\SI{0.907\pm0.017}{\giga\hertz}$.

\section{\label{sec:discussion}Discussion}

\begin{figure}[b!]
\centering
\includegraphics[width=0.9\columnwidth]{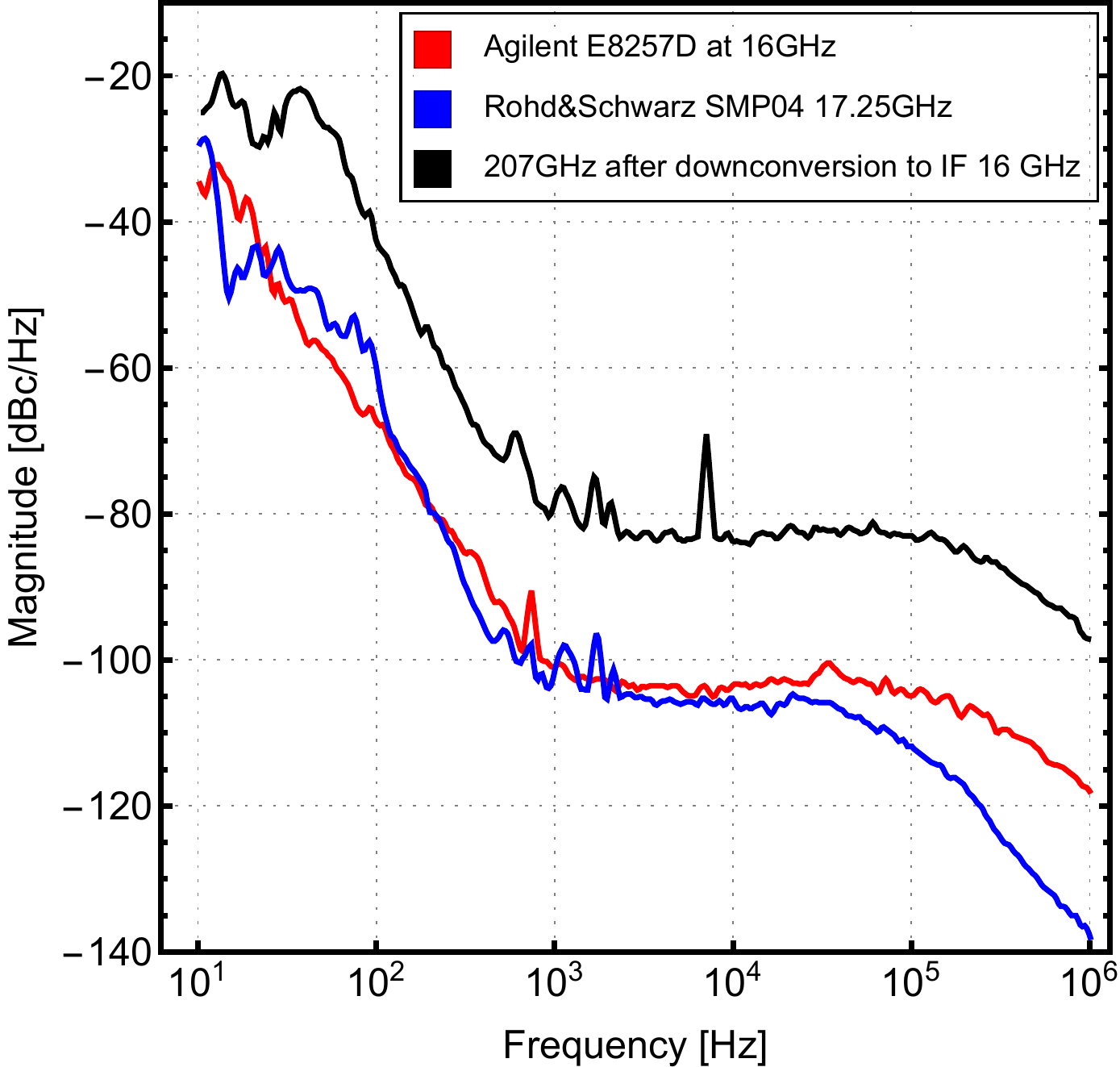}
\caption{Measured Phase noise. Black: \SI{207}{\giga\hertz} after downconversion to IF Comb, IF synthesizers and down-converted IF signal. Red: Agilent E8257D at \SI{16}{\giga\hertz}. Blue: Rohd\&Schwarz SMP04 at \SI{17.25}{\giga\hertz}. }
\label{fig:PhaseNoise}
\end{figure}

%\subsection{System Performance}
We can evaluate the performance of the experimental arrangement in terms of its application to spectroscopy. One of the key benefits of coherent THz generation is an improvement in the spectral purity of the resulting THz signal. In this experiment two low phase noise synthesizers were used for driving the optical frequency comb generator, and providing the LO for down-conversion. As discussed in the experimental methods section the resulting IF linewidth was measured and found to be \SI{900}{\hertz}. The phase noise of both synthesizers was measured together with the phase noise of the down-converted IF signal, shown in Figure \ref{fig:PhaseNoise}.
The down-converted IF signal was found to have phase noise \SI{26}{\dBc} higher than the phase noise of either of the two synthesisers. We attribute this to the frequency multiplication stage in the LO generation which adds a phase noise of $20 \log_{10}\left( M \right)$, where $M$ is the multiplication factor. We observe a Signal to Noise Ratio (SNR) of \SI{50}{\decibel} for the \SI{207}{\giga\hertz} signal down-converted to an IF of \SI{15}{\giga\hertz}.
Besides the frequency precision, the system is capable of high frequency accuracy, by referencing the microwave synthesizer to primary frequency standards via Global Positioning System (GPS).
Low power monochromatic illumination also allows minimum perturbation of the state of the sample, with negligible heating and perturbation of the population of the states.

%\subsection{Experimental limitations}

Some of the existing experimental limitations are imposed in a compromise for flexibility at the research and development stage and no attempts to improve acquisition rates were performed. %
The unpackaged UTC-PD required a fibre launcher to couple the focused laser beams into the device making it susceptible to drift. Continuous monitoring of the photocurrent allowed continuous optimisation of the alignment and normalisation of the observed curve for the small fluctuations observed. %

The output power available from UTC-PDs is limited primarily by thermal effects, while the bandwidth is limited by electron transit time and RC constraints. In addition to these bandwidth constraints, additional limitations are imposed on the device by the electronic circuitry connected to the device. For this experiment we utilised an UTC-PD integrated with co-planar waveguide. In order to extract the THz signals from the device an ACP probe with a WR-5 waveguide was used, along with a standard gain horn antenna. Such probes and antennas have a discrete bandwidth limitation imposed by the cut-off frequencies of the waveguide geometry. These limitations can be better managed through the use of devices integrated with planar broadband antennas, which remove the need for external probes to extract the THz signals\cite{Ito_2005,Rouvalis_2012}.

The main detection limitation was the requirement to multiply the LO signal for heterodyne detection. A possible improvement to the system would be to utilise the UTC-PD as a detector, or as a photonic LO for a Schottky barrier diode.
Utilising a self heterodyne configuration such as that of Hisatake \etal~\cite{Hisatake:2014} where one of the coherent lines is shifted in frequency and then used to down convert to a low frequency IF would enable the recovery of both amplitude and phase information from the spectrometer. In addition, it would eliminate linewidth broadening introduced by the LO multiplication step.

Acquisition rates are limited by the integration time in microwave spectrum analyser and the settling time of the waveshaper filter.

The reference signal for spectrum normalization neglects small changes in the sample reflections as a consequence of temperature variation of the sample refraction index.

%\subsection{\lihof measurements}
We performed all-optic very high resolution spectrometric measurements for \ho~to observe directly the octet hyperfine splitting of the $^5I_8$ ground state from the interaction with the \chem{Ho} nuclear angular momentum $I=\frac{7}{2}$. We observe a mean transition at \SI{205}{\giga\hertz} (\wn{6.8}) with hyperfine average spacing of \SI{4.4}{\giga\hertz}, consistent previous estimations~\cite{Ronnow2007} and with the latest revised calculations and assignments to the \ho electronic structure~\cite{Matmon2015}.

The spacing between adjacent hyperfine lines varies linearly through the octet as shown in Figure~\ref{fig:spacingtrend}, revealing the magnitude of the second order hyperfine shift (Eqn.10 in Matmon~\etal)~\cite{Matmon2015}. In previous measurements a second order term in the hyperfine lines spacing could appear experimentally as asymmetric line shape when the different hyperfine components overlap, but the high experimental resolution and low intrinsic linewidth in our system allows us to resolve the peaks differences individually.

Despite the extremely high resolution achieved in this experiment, it could be hard to directly discriminate between Gaussian, Lorentzian or Voight SLSs.  Nevertheless, the artificial broadening of the illumination SLS shown in Figure~\ref{fig:gallery} allowed us to observe a non-linear FWHM trend featuring a narrow \SI{99}{\percent} confidence band for the RMS fit. The trend in Figure~\ref{fig:broadening} is a clear signature of the characteristic of the convolution of two Gaussian SLSs. We attribute this signature to an intrinsically Gaussian absorbance line shape caused by a dominant inhomogeneous broadening. This allow us to rule out states lifetime as the main contributor to the absorbance linewidth and provides a lifetime lower boundary of \SI{1.103}{\nano\second}.

\section{\label{sec:conclusions}Conclusions}

We reported the development of a high resolution CW THz spectrometer enabled by ultra-fast photodiode technology and coherent photonic generation techniques. A spectral resolution of \SI{900}{\hertz} was achieved, allowing high resolution spectroscopy of \lihof at \SI{4}{\kelvin}.

We performed all-optic very high resolution spectrometric measurements for \ho that complement measurements by Matmon~\etal\cite{Matmon2015} for a band unreachable by FTIR. The spectral features of the transitions in the $^5I_8$ spin-orbit state have been resolved revealing the magnitude of the second order hyperfine splitting (\SI{-0.22\pm0.02}{\giga\hertz}) by direct optical observation.

In addition, we demonstrate a novel software defined spectroscopy concept that digitally enhances the performance of our high resolution CW THz spectrometer. We exploit that enhancement to confirm the absorbance spectral line shapes indirectly, not by their apparent shape but by analysing the FWHM trend. This allowed us to determine that the absorbance spectral line shape is predominantly Gaussian; therefore by attributing the broadening to an inhomogeneous process, we can conclude that the intrinsic state lifetimes are much longer than the inverse line-width. $\tau \gg \SI{1.1}{\nano\second}$ ($1/\SI{0.907}{\giga\hertz} $).

\section{Acknowledgements}
The authors acknowledge financial support from EPSRC grants EP/J017671/1 ``Coherent Terahertz Systems'' and EP/P021859/1 ``HyperTerahertz - High precision terahertz spectroscopy and microscopy''.
RIH and HS thank Guy Matmon and Joshua Freeman for valuable discussion.
RIH, HS, AJS and GA declare financial interest relating to a patent application for the concept of software defined spectroscopy.

\section{Author information}

%\subsection{Affiliations}
%
%EEE, UCL:
%RIH, HS, SPS, AJS
%
%LCN, UCL:
%RIH, AJS
%
%Physics, ETH Z\"{u}rich:
%GA
%
%PSI:
%GA

\subsection{Contributions}

GA and AJS
conceived the experiment. %
RIH, HS, GA and AJS
conceived and developed the idea of software defined spectroscopy. %
HS %
implemented the photonics system.
RIH %
implemented sample and cryogenics system.
RIH and HS %
collected and analysed data.
RIH %
developed the analytical description of Markov OOK and SLS discrimination..
JPS %
analyzed and quantified the system performance.
RIH and JPS %
wrote the manuscript.
All authors %
contributed to manuscript revision and discussed the results.
GA and AJS %
supervised the work.

\subsection{Corresponding authors}

Correspondence to RIH (r.hermans@ucl.ac.uk).

\section{References}

%GATHER{C:\Users\Rodolfo\Docs\BibliographicReferences\COTS.bib}
%\bibliography{COTS}
%\clearpage
%\newpage

\clearpage
\newpage
\appendix
\section{On-Off Keying Stochastic Modulation}

Let's assume two signals at the UTC-PD with arbitrary amplitudes and relative phase $\phi$

%\subsection{no modulation}
%\begin{eqnarray}\label{RHN1}
%% \nonumber to remove numbering (before each equation)
%    \mathcal{A}_1(t) &=& a_1 \, \exp{ \left(i (\omega_1 t)\right)} \nonumber\\
%    \mathcal{A}_2(t) &=& a_2 \, \exp{ \left(i (\omega_2 t + \phi)\right)} \nonumber
%\end{eqnarray}
%
%The prosthaphaeresis reverse trigonometric identities  implies that the total $\mathcal{A}(t)=\mathcal{A}_1(t)+\mathcal{A}_2(t)$
%
%\begin{eqnarray}\label{RHN1}
%% \nonumber to remove numbering (before each equation)
%    \mathcal{A}(t)&=& (a_1 - a_2) \, \exp{ \left(i (\omega_1 t)\right)}\nonumber\\
%          & & +  a_2 ( \exp{ \left(i (\omega_1 t)\right)}+ \exp{ \left(i (\omega_2 t + \phi)\right)} )\nonumber\\
%    &=& (a_1 - a_2) \, \exp{ \left(i (\omega_1 t)\right)} \nonumber\\
%    & & + 2 a_2 \left( \exp{ \left(i( \frac{\omega_1 + \omega_2}{2} t + \phi )\right)} \exp{ \left(i(  \frac{\omega_1 - \omega_2}{2} t - \phi)\right)}\right)\nonumber\\
%\end{eqnarray}
%
%If the two components have identical amplitude ($(a = a_1 = a_2)$) then we have only a beating signal
%The photocurrent in the UTC-PD is proportional to the intensity $ \mathcal{I}(t)= \left| \mathcal{A}(t) \right|^2$ and therefore oscillating with frequency $\omega_1 - \omega_2$.

\subsection{General modulation}

If one component is modulated

\begin{eqnarray}\label{RHN1}
    \mathcal{A}_1(t) &=& a_1 \, \exp{ \left(i (\omega_1 t)\right)} \nonumber\\
    \mathcal{A}_2(t) &=& a_2(t) \, \exp{ \left(i (\omega_2 t + \phi)\right)} \nonumber
\end{eqnarray}

The prosthaphaeresis reverse trigonometric identities  implies that the total $\mathcal{A}(t)=\mathcal{A}_1(t)+\mathcal{A}_2(t)$

\begin{eqnarray}\label{RHN2}
    \mathcal{A}(t)&=& (a_1 - a_2(t)) \, \exp{ \left(i (\omega_1 t)\right)}\nonumber\\
          & & +  a_2(t) ( \exp{ \left(i (\omega_1 t)\right)}+ \exp{ \left(i (\omega_2 t + \phi)\right)} )\nonumber\\
    &=& (a_1 - a_2(t)) \, \exp{ \left(i (\omega_1 t)\right)} \nonumber\\
    & & + 2 a_2(t)  \exp{ \left(i( \frac{\omega_1 + \omega_2}{2} t + \phi )\right)} \nonumber\\
    & & \times \exp{ \left(i(  \frac{\omega_1 - \omega_2}{2} t - \phi)\right)} \nonumber\\
\end{eqnarray}

The photocurrent in the UTC-PD is proportional to the intensity $ \mathcal{I}(t) \propto \left| \mathcal{A}(t) \right|^2$, the high-frequency are filtered by the limited carrier speed and the low frequency current oscillation is  \mbox{$\Delta\omega = \omega_1 - \omega_2$}.

\begin{eqnarray}\label{RHN3}
 \mathcal{I}(\omega) &\propto &  a_2(\omega) \otimes \delta\left(\omega - \Delta\omega\right)
\end{eqnarray}

\subsection{Stochastic OOK}
We are concerned with the case where $ a_2(t)$ is either $1$ or $0$ representing the simplest form of amplitude-shift keying, the On-off keying (OOK). The transitions are defined  by a two state Markov chain stochastic process, describing a sequence of possible On/Off events in which the probability of each event depends only on the previous state and a constant matrix $m$.

\begin{equation}\label{eq:stocmatrixSM}
m = \left(
\begin{array}{cc}
 p & 1-p \\
 1-p & p \\
\end{array}
\right)
\end{equation}

\begin{figure}[tb!]
\vskip 5mm
\begin{subfigure}{\columnwidth}
\includegraphics[width=\columnwidth]{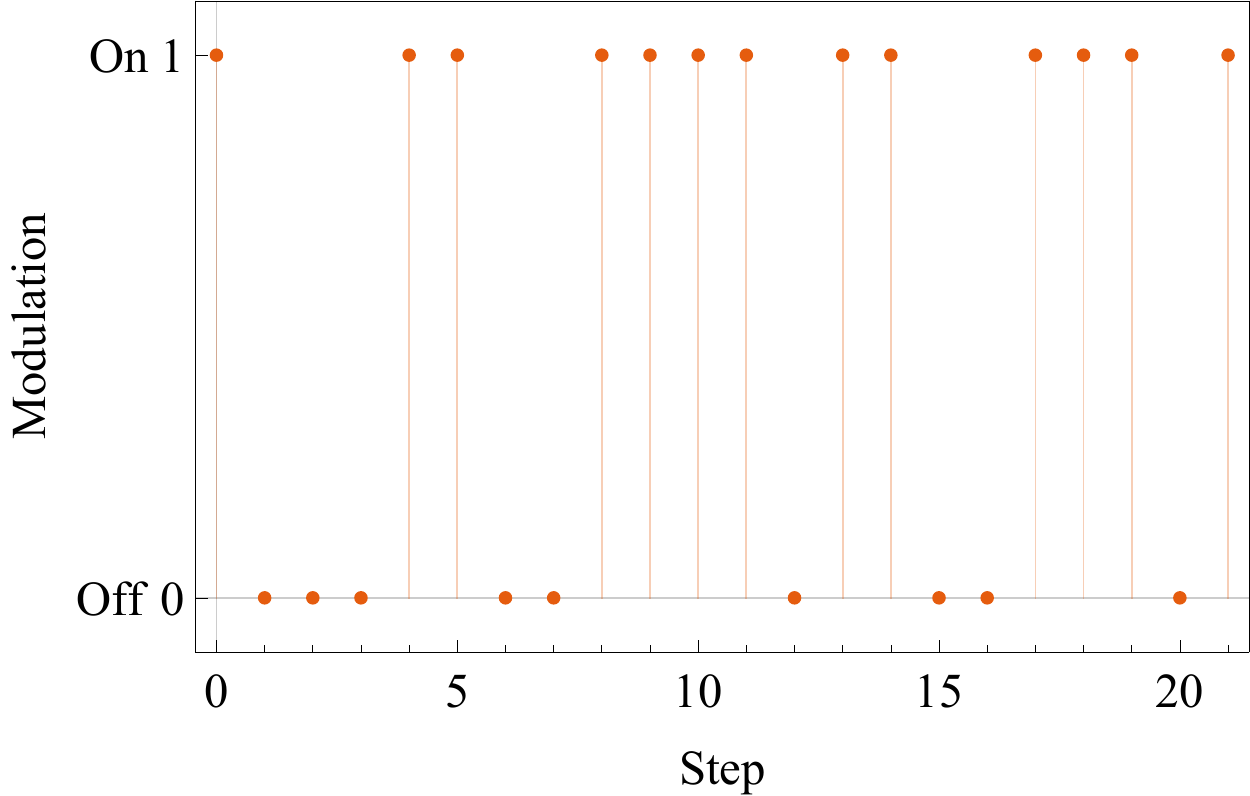}
\caption{Discrete case, a chain of \emph{on} (1) \emph{off} (0) states stochastically defined in a chain. }
\end{subfigure}\\
\vskip 5mm
\begin{subfigure}{\columnwidth}
\includegraphics[width=\columnwidth]{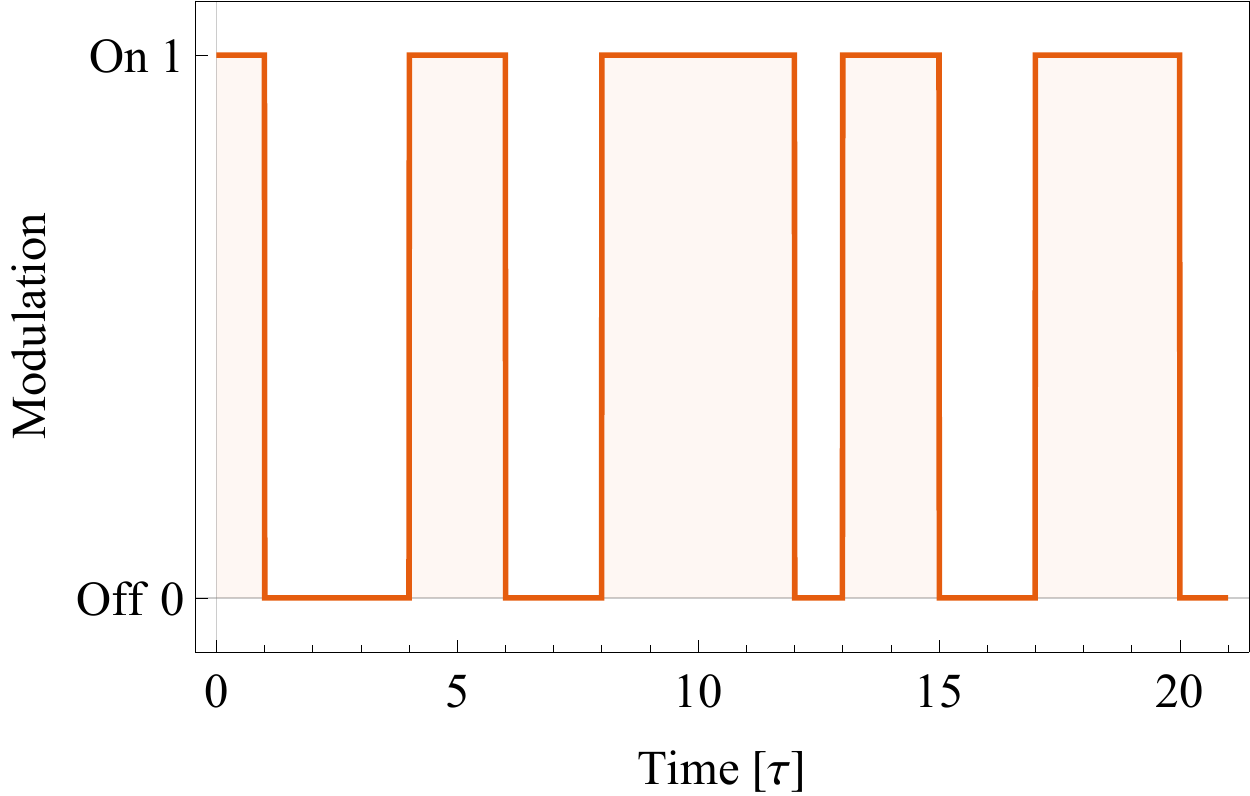}
\caption{Continuos time case, \emph{on} (1) and \emph{off} states are defined for a continuos time with constant value between transitions that occur only at integer values of $tau$.}
\end{subfigure}
\caption{Modulation as a stochastic process}
\label{fig:discretemarkovchain}

\end{figure}

Assuming that we start in the state "on", $s=\{1,0\}$ a discrete Markov chain could look like in Figure~\ref{fig:discretemarkovchain}a

the probability density function for the $n^{\textrm{th}}$ state is given by

\begin{equation}\label{eq:pdf}
 P(\text{"On"}|n) =  \frac{1}{2} \left(1-(2 p-1)^n\right)
\end{equation}

The modulation spectra $a_2(\omega)$ can be obtained from the Wiener-Khinchin-Einstein theorem~\cite{Khintchine1934} by first determining autocorrelation function. Based on the pdf function in Eqn~\ref{eq:pdf} we can show that the autocorrelation function for the discrete case is given by \mbox{equation~(\ref{eq:autocorrmk})}.
\begin{equation}\label{eq:autocorrmk}
 C(n) = \begin{cases}
   (2 p -1)^{n} & p \neq 1/2 \\ %\, \wedge\, 0<p<1  \\
    \delta_{n\,0} & p=1/2
 \end{cases}
\end{equation}

Three distinguishable cases exist for the correlation function. For $p=1/2$ corresponds to an unbiased random bit sequence where each bit value is independent and the correlation is a $ \delta_{i\,0}$. For $1/2 < p < 1$ the correlation decays exponentially. And finally the case $0 < p < 1/2$ the correlation function is oscillatory with exponentially decaying amplitude.

In the continuous case each step in the chain represents a constant value over a time $\tau$, and the Markov chain corresponds to a trend of square pulses, and the correlation function is the linear interpolation between the values in \mbox{equation~(\ref{eq:autocorrmk})} separated by time $\tau$.

\begin{figure}[tb!]
\vskip 5mm
\begin{subfigure}{\columnwidth}
\includegraphics[width=\columnwidth]{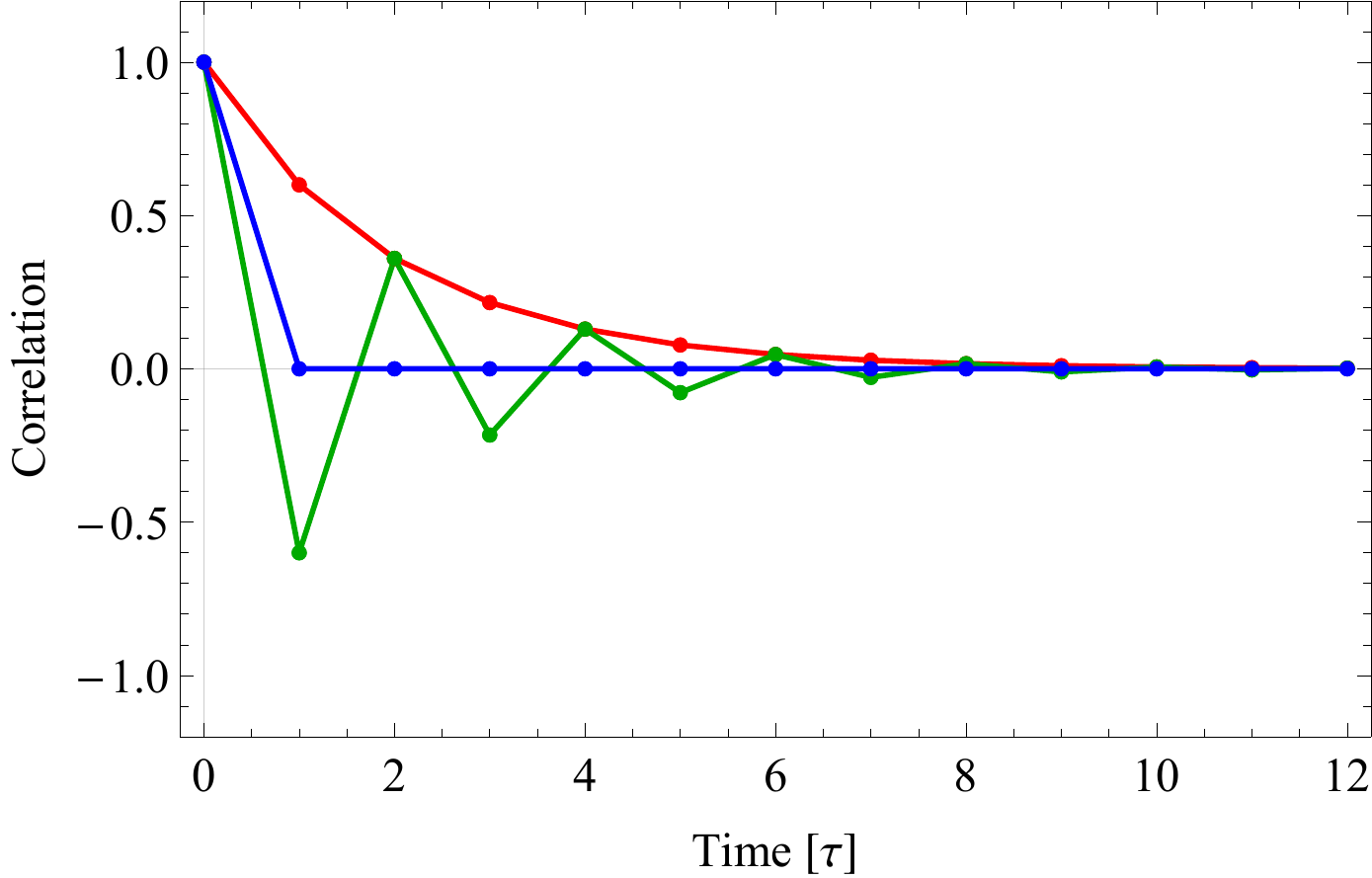}
\caption{Correlation function for $p=0.2$,  $p=0.5$ and $p=0.8$}
\end{subfigure}
\vskip 5mm
\begin{subfigure}{\columnwidth}
\includegraphics[width=\columnwidth]{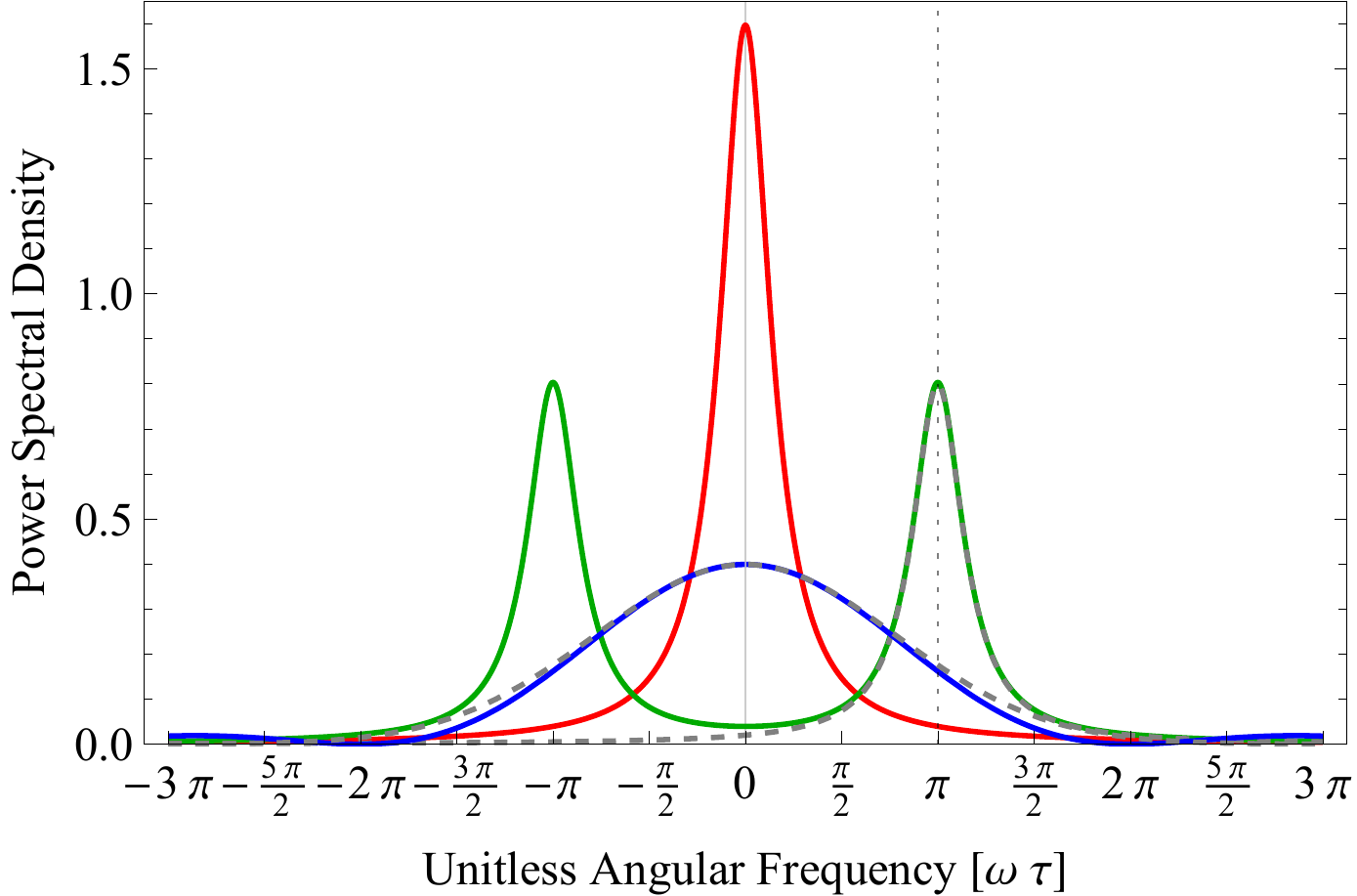}
\caption{Continuos case}
\end{subfigure}
\caption{Power Spectral Density for  $p=0.2$,  $p=0.5$ and $p=0.8$. Dashed gray lines correspond to the approximation in \mbox{equation~(\ref{eq:contSpectAppr})}}
\label{fig:CorrAndSpectr}
\end{figure}

Figure~\ref{fig:CorrAndSpectr}a shows three example correlation function, an oscillatory system with $p=0.2$ (green) and a exponentially decaying system with $p=0.8$ and an uncorrelated system with $p=0.5$.

The continuous case can be written as

\begin{equation}\label{eq:contCorr}
C(t) = \begin{cases}
 e^{-\alpha t} \cos (2 \pi  f t) & 0<p<\frac{1}{2} \,\wedge\, t>0      \\
 e^{-\alpha t}                   & \frac{1}{2}<p<1 \,\wedge\, t>0      \\
 (\tau - t)/\tau^2               & p = \frac{1}{2} \,\wedge\, 0<t<\tau \\
 0                               & p = \frac{1}{2} \,\wedge\, t>\tau
\end{cases}
\end{equation}

where $\alpha = -\log (\left|2 p-1\right|) \tau^{-1} $ and $f=(2 \tau)^{-1}$. Now the Wiener-Khinchin-Einstein theorem\cite{Khintchine1934} implies that the power spectral density of the system $S(\omega)$ is given by the inverse Fourier transform of the correlation function. The results are shown in Figure~\ref{fig:CorrAndSpectr}b. Therefore

\begin{equation}\label{eq:contSpectExact}
S(\omega) = \begin{cases}
\sqrt{\frac{2}{\pi }} \frac{ \alpha  \tau ^2 \left(\tau ^2 \left(\alpha ^2+\omega ^2\right)+\pi ^2\right)}{\tau ^4 \left(\alpha ^2+\omega ^2\right)^2+2 \pi ^2 \tau ^2 (\alpha -\omega ) (\alpha +\omega )+\pi ^4} & 0<p<\frac{1}{2}      \\ \\
\frac{\sqrt{\frac{2}{\pi }} \alpha }{\alpha ^2+\omega ^2}               & \frac{1}{2}<p<1     \\ \\
\frac{\sqrt{\frac{2}{\pi }} (1-\cos (\tau  \omega ))}{\tau ^2 \omega ^2}              & p = \frac{1}{2} \\
\end{cases}
\end{equation}

We recognize the exact solution for the case $\frac{1}{2}<p<1$ as a Lorentzian. The case $0<p<\frac{1}{2}$ features a side peak under the condition $\alpha  \tau <\sqrt{3} \pi$

The other two cases to good approximation can be described by

\begin{equation}\label{eq:contSpectAppr}
S(\omega) \approx \begin{cases}
 \frac{c\, \gamma }{2 \pi  \left(\frac{\gamma ^2}{4}+\left(\omega -\frac{\pi }{\tau }\right)^2\right)} & 0<p<\frac{1}{2}  \\ \\
\frac{1}{\sqrt{2 \pi }}\,e^{-\frac{1}{2} \left(\frac{\tau  \omega}{ \sqrt{6}}\right) ^2}                                           & p = \frac{1}{2}  \\
\end{cases}
\end{equation}

where $\gamma = 2 \sqrt{2 \tau} \pi  \alpha  / \sqrt{2 \pi ^2-\alpha ^2 \tau ^2}$ and $c=\frac{\sqrt{\pi } \sqrt{\tau }}{2 \sqrt{2 \pi ^2-\alpha ^2 \tau ^2}}$. The Approximations are Taylor series up to 2nd order for $\omega\approx\pi/\tau$ and $\alpha \approx 0$ for $0<p<\frac{1}{2}$ and $\omega\approx 0$ for  $p = \frac{1}{2}$.

%\bibliography{COTS}

\begin{thebibliography}{52}%
\makeatletter
\providecommand \@ifxundefined [1]{%
 \@ifx{#1\undefined}
}%
\providecommand \@ifnum [1]{%
 \ifnum #1\expandafter \@firstoftwo
 \else \expandafter \@secondoftwo
 \fi
}%
\providecommand \@ifx [1]{%
 \ifx #1\expandafter \@firstoftwo
 \else \expandafter \@secondoftwo
 \fi
}%
\providecommand \natexlab [1]{#1}%
\providecommand \enquote  [1]{``#1''}%
\providecommand \bibnamefont  [1]{#1}%
\providecommand \bibfnamefont [1]{#1}%
\providecommand \citenamefont [1]{#1}%
\providecommand \href@noop [0]{\@secondoftwo}%
\providecommand \href [0]{\begingroup \@sanitize@url \@href}%
\providecommand \@href[1]{\@@startlink{#1}\@@href}%
\providecommand \@@href[1]{\endgroup#1\@@endlink}%
\providecommand \@sanitize@url [0]{\catcode `\\12\catcode `\$12\catcode
  `\&12\catcode `\#12\catcode `\^12\catcode `\_12\catcode `\%12\relax}%
\providecommand \@@startlink[1]{}%
\providecommand \@@endlink[0]{}%
\providecommand \url  [0]{\begingroup\@sanitize@url \@url }%
\providecommand \@url [1]{\endgroup\@href {#1}{\urlprefix }}%
\providecommand \urlprefix  [0]{URL }%
\providecommand \Eprint [0]{\href }%
\providecommand \doibase [0]{http://dx.doi.org/}%
\providecommand \selectlanguage [0]{\@gobble}%
\providecommand \bibinfo  [0]{\@secondoftwo}%
\providecommand \bibfield  [0]{\@secondoftwo}%
\providecommand \translation [1]{[#1]}%
\providecommand \BibitemOpen [0]{}%
\providecommand \bibitemStop [0]{}%
\providecommand \bibitemNoStop [0]{.\EOS\space}%
\providecommand \EOS [0]{\spacefactor3000\relax}%
\providecommand \BibitemShut  [1]{\csname bibitem#1\endcsname}%
\let\auto@bib@innerbib\@empty
%</preamble>
\bibitem [{\citenamefont {Dhillon}\ \emph {et~al.}(2017)\citenamefont
  {Dhillon}, \citenamefont {Vitiello}, \citenamefont {Linfield}, \citenamefont
  {Davies}, \citenamefont {Hoffmann}, \citenamefont {Booske}, \citenamefont
  {Paoloni}, \citenamefont {Gensch}, \citenamefont {Weightman}, \citenamefont
  {Williams}, \citenamefont {Castro-Camus}, \citenamefont {Cumming},
  \citenamefont {Simoens}, \citenamefont {Escorcia-Carranza}, \citenamefont
  {Grant}, \citenamefont {Lucyszyn}, \citenamefont {Kuwata-Gonokami},
  \citenamefont {Konishi}, \citenamefont {Koch}, \citenamefont {Schmuttenmaer},
  \citenamefont {Cocker}, \citenamefont {Huber}, \citenamefont {Markelz},
  \citenamefont {Taylor}, \citenamefont {Wallace}, \citenamefont {Zeitler},
  \citenamefont {Sibik}, \citenamefont {Korter}, \citenamefont {Ellison},
  \citenamefont {Rea}, \citenamefont {Goldsmith}, \citenamefont {Cooper},
  \citenamefont {Appleby}, \citenamefont {Pardo}, \citenamefont {Huggard},
  \citenamefont {Krozer}, \citenamefont {Shams}, \citenamefont {Fice},
  \citenamefont {Renaud}, \citenamefont {Seeds}, \citenamefont {St{\"{o}}hr},
  \citenamefont {Naftaly}, \citenamefont {Ridler}, \citenamefont {Clarke},
  \citenamefont {Cunningham},\ and\ \citenamefont {Johnston}}]{Dhillon_2017}%
  \BibitemOpen
  \bibfield  {author} {\bibinfo {author} {\bibfnamefont {S.~S.}\ \bibnamefont
  {Dhillon}}, \bibinfo {author} {\bibfnamefont {M.~S.}\ \bibnamefont
  {Vitiello}}, \bibinfo {author} {\bibfnamefont {E.~H.}\ \bibnamefont
  {Linfield}}, \bibinfo {author} {\bibfnamefont {A.~G.}\ \bibnamefont
  {Davies}}, \bibinfo {author} {\bibfnamefont {M.~C.}\ \bibnamefont
  {Hoffmann}}, \bibinfo {author} {\bibfnamefont {J.}~\bibnamefont {Booske}},
  \bibinfo {author} {\bibfnamefont {C.}~\bibnamefont {Paoloni}}, \bibinfo
  {author} {\bibfnamefont {M.}~\bibnamefont {Gensch}}, \bibinfo {author}
  {\bibfnamefont {P.}~\bibnamefont {Weightman}}, \bibinfo {author}
  {\bibfnamefont {G.~P.}\ \bibnamefont {Williams}}, \bibinfo {author}
  {\bibfnamefont {E.}~\bibnamefont {Castro-Camus}}, \bibinfo {author}
  {\bibfnamefont {D.~R.~S.}\ \bibnamefont {Cumming}}, \bibinfo {author}
  {\bibfnamefont {F.}~\bibnamefont {Simoens}}, \bibinfo {author} {\bibfnamefont
  {I.}~\bibnamefont {Escorcia-Carranza}}, \bibinfo {author} {\bibfnamefont
  {J.}~\bibnamefont {Grant}}, \bibinfo {author} {\bibfnamefont
  {S.}~\bibnamefont {Lucyszyn}}, \bibinfo {author} {\bibfnamefont
  {M.}~\bibnamefont {Kuwata-Gonokami}}, \bibinfo {author} {\bibfnamefont
  {K.}~\bibnamefont {Konishi}}, \bibinfo {author} {\bibfnamefont
  {M.}~\bibnamefont {Koch}}, \bibinfo {author} {\bibfnamefont {C.~A.}\
  \bibnamefont {Schmuttenmaer}}, \bibinfo {author} {\bibfnamefont {T.~L.}\
  \bibnamefont {Cocker}}, \bibinfo {author} {\bibfnamefont {R.}~\bibnamefont
  {Huber}}, \bibinfo {author} {\bibfnamefont {A.~G.}\ \bibnamefont {Markelz}},
  \bibinfo {author} {\bibfnamefont {Z.~D.}\ \bibnamefont {Taylor}}, \bibinfo
  {author} {\bibfnamefont {V.~P.}\ \bibnamefont {Wallace}}, \bibinfo {author}
  {\bibfnamefont {J.~A.}\ \bibnamefont {Zeitler}}, \bibinfo {author}
  {\bibfnamefont {J.}~\bibnamefont {Sibik}}, \bibinfo {author} {\bibfnamefont
  {T.~M.}\ \bibnamefont {Korter}}, \bibinfo {author} {\bibfnamefont
  {B.}~\bibnamefont {Ellison}}, \bibinfo {author} {\bibfnamefont
  {S.}~\bibnamefont {Rea}}, \bibinfo {author} {\bibfnamefont {P.}~\bibnamefont
  {Goldsmith}}, \bibinfo {author} {\bibfnamefont {K.~B.}\ \bibnamefont
  {Cooper}}, \bibinfo {author} {\bibfnamefont {R.}~\bibnamefont {Appleby}},
  \bibinfo {author} {\bibfnamefont {D.}~\bibnamefont {Pardo}}, \bibinfo
  {author} {\bibfnamefont {P.~G.}\ \bibnamefont {Huggard}}, \bibinfo {author}
  {\bibfnamefont {V.}~\bibnamefont {Krozer}}, \bibinfo {author} {\bibfnamefont
  {H.}~\bibnamefont {Shams}}, \bibinfo {author} {\bibfnamefont
  {M.}~\bibnamefont {Fice}}, \bibinfo {author} {\bibfnamefont {C.}~\bibnamefont
  {Renaud}}, \bibinfo {author} {\bibfnamefont {A.}~\bibnamefont {Seeds}},
  \bibinfo {author} {\bibfnamefont {A.}~\bibnamefont {St{\"{o}}hr}}, \bibinfo
  {author} {\bibfnamefont {M.}~\bibnamefont {Naftaly}}, \bibinfo {author}
  {\bibfnamefont {N.}~\bibnamefont {Ridler}}, \bibinfo {author} {\bibfnamefont
  {R.}~\bibnamefont {Clarke}}, \bibinfo {author} {\bibfnamefont {J.~E.}\
  \bibnamefont {Cunningham}}, \ and\ \bibinfo {author} {\bibfnamefont {M.~B.}\
  \bibnamefont {Johnston}},\ }\bibfield  {title} {\enquote {\bibinfo {title}
  {{The 2017 terahertz science and technology roadmap}},}\ }\href {\doibase
  10.1088/1361-6463/50/4/043001} {\bibfield  {journal} {\bibinfo  {journal}
  {Journal of Physics D: Applied Physics}\ }\textbf {\bibinfo {volume} {50}},\
  \bibinfo {pages} {43001} (\bibinfo {year} {2017})}\BibitemShut {NoStop}%
\bibitem [{\citenamefont {Seeds}\ \emph {et~al.}(2013)\citenamefont {Seeds},
  \citenamefont {Fice}, \citenamefont {Balakier}, \citenamefont {Natrella},
  \citenamefont {Mitrofanov}, \citenamefont {Lamponi}, \citenamefont {Chtioui},
  \citenamefont {van Dijk}, \citenamefont {Pepper}, \citenamefont {Aeppli},
  \citenamefont {Davies}, \citenamefont {Dean}, \citenamefont {Linfield},\ and\
  \citenamefont {Renaud}}]{Seeds2013}%
  \BibitemOpen
  \bibfield  {author} {\bibinfo {author} {\bibfnamefont {A.~J.}\ \bibnamefont
  {Seeds}}, \bibinfo {author} {\bibfnamefont {M.~J.}\ \bibnamefont {Fice}},
  \bibinfo {author} {\bibfnamefont {K.}~\bibnamefont {Balakier}}, \bibinfo
  {author} {\bibfnamefont {M.}~\bibnamefont {Natrella}}, \bibinfo {author}
  {\bibfnamefont {O.}~\bibnamefont {Mitrofanov}}, \bibinfo {author}
  {\bibfnamefont {M.}~\bibnamefont {Lamponi}}, \bibinfo {author} {\bibfnamefont
  {M.}~\bibnamefont {Chtioui}}, \bibinfo {author} {\bibfnamefont
  {F.}~\bibnamefont {van Dijk}}, \bibinfo {author} {\bibfnamefont
  {M.}~\bibnamefont {Pepper}}, \bibinfo {author} {\bibfnamefont
  {G.}~\bibnamefont {Aeppli}}, \bibinfo {author} {\bibfnamefont {A.~G.}\
  \bibnamefont {Davies}}, \bibinfo {author} {\bibfnamefont {P.}~\bibnamefont
  {Dean}}, \bibinfo {author} {\bibfnamefont {E.}~\bibnamefont {Linfield}}, \
  and\ \bibinfo {author} {\bibfnamefont {C.~C.}\ \bibnamefont {Renaud}},\
  }\bibfield  {title} {{\selectlanguage {English}\enquote {\bibinfo {title}
  {{Coherent terahertz photonics}},}\ }}\href {\doibase 10.1364/oe.21.022988}
  {\bibfield  {journal} {\bibinfo  {journal} {Optics Express}\ }\textbf
  {\bibinfo {volume} {21}},\ \bibinfo {pages} {22988} (\bibinfo {year}
  {2013})}\BibitemShut {NoStop}%
\bibitem [{\citenamefont {Hintzsche}\ and\ \citenamefont
  {Stopper}(2012)}]{Hintzsche2012}%
  \BibitemOpen
  \bibfield  {author} {\bibinfo {author} {\bibfnamefont {H.}~\bibnamefont
  {Hintzsche}}\ and\ \bibinfo {author} {\bibfnamefont {H.}~\bibnamefont
  {Stopper}},\ }\href {\doibase 10.1080/10643389.2011.574206} {\enquote
  {\bibinfo {title} {{Effects of terahertz radiation on biological systems}},}\
  } (\bibinfo {year} {2012})\BibitemShut {NoStop}%
\bibitem [{\citenamefont {Graf}\ \emph {et~al.}(2015)\citenamefont {Graf},
  \citenamefont {Honingh}, \citenamefont {Jacobs},\ and\ \citenamefont
  {Stutzki}}]{Graf2015}%
  \BibitemOpen
  \bibfield  {author} {\bibinfo {author} {\bibfnamefont {U.~U.}\ \bibnamefont
  {Graf}}, \bibinfo {author} {\bibfnamefont {C.~E.}\ \bibnamefont {Honingh}},
  \bibinfo {author} {\bibfnamefont {K.}~\bibnamefont {Jacobs}}, \ and\ \bibinfo
  {author} {\bibfnamefont {J.}~\bibnamefont {Stutzki}},\ }\bibfield  {title}
  {\enquote {\bibinfo {title} {{Terahertz Heterodyne Array Receivers for
  Astronomy}},}\ }\href {\doibase 10.1007/s10762-015-0171-7} {\bibfield
  {journal} {\bibinfo  {journal} {Journal of Infrared, Millimeter, and
  Terahertz Waves}\ }\textbf {\bibinfo {volume} {36}},\ \bibinfo {pages}
  {896--921} (\bibinfo {year} {2015})}\BibitemShut {NoStop}%
\bibitem [{\citenamefont {Seeds}\ \emph {et~al.}(2015)\citenamefont {Seeds},
  \citenamefont {Shams}, \citenamefont {Fice},\ and\ \citenamefont
  {Renaud}}]{Seeds2015}%
  \BibitemOpen
  \bibfield  {author} {\bibinfo {author} {\bibfnamefont {A.~J.}\ \bibnamefont
  {Seeds}}, \bibinfo {author} {\bibfnamefont {H.}~\bibnamefont {Shams}},
  \bibinfo {author} {\bibfnamefont {M.~J.}\ \bibnamefont {Fice}}, \ and\
  \bibinfo {author} {\bibfnamefont {C.~C.}\ \bibnamefont {Renaud}},\ }\bibfield
   {title} {\enquote {\bibinfo {title} {{TeraHertz Photonics for Wireless
  Communications}},}\ }\href {\doibase 10.1109/JLT.2014.2355137} {\bibfield
  {journal} {\bibinfo  {journal} {Journal of Lightwave Technology}\ }\textbf
  {\bibinfo {volume} {33}},\ \bibinfo {pages} {579--587} (\bibinfo {year}
  {2015})}\BibitemShut {NoStop}%
\bibitem [{\citenamefont {Stuart}(2015)}]{IRSpecRev}%
  \BibitemOpen
  \bibfield  {author} {\bibinfo {author} {\bibfnamefont {B.}~\bibnamefont
  {Stuart}},\ }\enquote {\bibinfo {title} {{Infrared Spectroscopy}},}\ in\
  \href {\doibase 10.1002/0471238961.0914061810151405.a01.pub3} {\emph
  {\bibinfo {booktitle} {Kirk‐Othmer Encyclopedia of Chemical Technology}}}\
  (\bibinfo  {publisher} {American Cancer Society},\ \bibinfo {year} {2015})\
  pp.\ \bibinfo {pages} {1--18}\BibitemShut {NoStop}%
\bibitem [{\citenamefont {Bruker}(2018)}]{BrukerWebIF125SHR}%
  \BibitemOpen
  \bibfield  {author} {\bibinfo {author} {\bibnamefont {Bruker}},\ }\href
  {https://www.bruker.com/products/infrared-near-infrared-and-raman-spectroscopy/ft-ir-research-spectrometers/ifs-125hr/overview.html}
  {\enquote {\bibinfo {title} {{The IFS 125HR FTIR Spectrometer}},}\ }
  (\bibinfo {year} {2018})\BibitemShut {NoStop}%
\bibitem [{\citenamefont {Elzinga}\ \emph {et~al.}(1987)\citenamefont
  {Elzinga}, \citenamefont {Kneisler}, \citenamefont {Lytle}, \citenamefont
  {Jiang}, \citenamefont {King},\ and\ \citenamefont
  {Laurendeau}}]{Elzinga1987}%
  \BibitemOpen
  \bibfield  {author} {\bibinfo {author} {\bibfnamefont {P.~A.}\ \bibnamefont
  {Elzinga}}, \bibinfo {author} {\bibfnamefont {R.~J.}\ \bibnamefont
  {Kneisler}}, \bibinfo {author} {\bibfnamefont {F.~E.}\ \bibnamefont {Lytle}},
  \bibinfo {author} {\bibfnamefont {Y.}~\bibnamefont {Jiang}}, \bibinfo
  {author} {\bibfnamefont {G.~B.}\ \bibnamefont {King}}, \ and\ \bibinfo
  {author} {\bibfnamefont {N.~M.}\ \bibnamefont {Laurendeau}},\ }\bibfield
  {title} {\enquote {\bibinfo {title} {{Pump/probe method for fast analysis of
  visible spectral signatures utilizing asynchronous optical sampling}},}\
  }\href {\doibase 10.1364/AO.26.004303} {\bibfield  {journal} {\bibinfo
  {journal} {Applied Optics}\ }\textbf {\bibinfo {volume} {26}},\ \bibinfo
  {pages} {4303} (\bibinfo {year} {1987})}\BibitemShut {NoStop}%
\bibitem [{\citenamefont {Bartels}\ \emph {et~al.}(2007)\citenamefont
  {Bartels}, \citenamefont {Cerna}, \citenamefont {Kistner}, \citenamefont
  {Thoma}, \citenamefont {Hudert}, \citenamefont {Janke},\ and\ \citenamefont
  {Dekorsy}}]{Bartels2007}%
  \BibitemOpen
  \bibfield  {author} {\bibinfo {author} {\bibfnamefont {A.}~\bibnamefont
  {Bartels}}, \bibinfo {author} {\bibfnamefont {R.}~\bibnamefont {Cerna}},
  \bibinfo {author} {\bibfnamefont {C.}~\bibnamefont {Kistner}}, \bibinfo
  {author} {\bibfnamefont {A.}~\bibnamefont {Thoma}}, \bibinfo {author}
  {\bibfnamefont {F.}~\bibnamefont {Hudert}}, \bibinfo {author} {\bibfnamefont
  {C.}~\bibnamefont {Janke}}, \ and\ \bibinfo {author} {\bibfnamefont
  {T.}~\bibnamefont {Dekorsy}},\ }\bibfield  {title} {\enquote {\bibinfo
  {title} {{Ultrafast time-domain spectroscopy based on high-speed asynchronous
  optical sampling}},}\ }\href {\doibase 10.1063/1.2714048} {\bibfield
  {journal} {\bibinfo  {journal} {Review of Scientific Instruments}\ }\textbf
  {\bibinfo {volume} {78}},\ \bibinfo {pages} {035107} (\bibinfo {year}
  {2007})}\BibitemShut {NoStop}%
\bibitem [{\citenamefont {Stanze}\ \emph {et~al.}(2011)\citenamefont {Stanze},
  \citenamefont {Deninger}, \citenamefont {Roggenbuck}, \citenamefont
  {Schindler}, \citenamefont {Schlak},\ and\ \citenamefont
  {Sartorius}}]{Stanze2011}%
  \BibitemOpen
  \bibfield  {author} {\bibinfo {author} {\bibfnamefont {D.}~\bibnamefont
  {Stanze}}, \bibinfo {author} {\bibfnamefont {A.}~\bibnamefont {Deninger}},
  \bibinfo {author} {\bibfnamefont {A.}~\bibnamefont {Roggenbuck}}, \bibinfo
  {author} {\bibfnamefont {S.}~\bibnamefont {Schindler}}, \bibinfo {author}
  {\bibfnamefont {M.}~\bibnamefont {Schlak}}, \ and\ \bibinfo {author}
  {\bibfnamefont {B.}~\bibnamefont {Sartorius}},\ }\bibfield  {title} {\enquote
  {\bibinfo {title} {{Compact cw Terahertz Spectrometer Pumped at 1.5 $\mu$m
  Wavelength}},}\ }\href {\doibase 10.1007/s10762-010-9751-8} {\bibfield
  {journal} {\bibinfo  {journal} {Journal of Infrared, Millimeter, and
  Terahertz Waves}\ }\textbf {\bibinfo {volume} {32}},\ \bibinfo {pages}
  {225--232} (\bibinfo {year} {2011})}\BibitemShut {NoStop}%
\bibitem [{\citenamefont {{Balakier}}\ \emph {et~al.}(2018)\citenamefont
  {{Balakier}}, \citenamefont {{Shams}}, \citenamefont {{Fice}}, \citenamefont
  {{Ponnampalam}}, \citenamefont {{Graham}}, \citenamefont {{Renaud}},\ and\
  \citenamefont {{Seeds}}}]{8388199}%
  \BibitemOpen
  \bibfield  {author} {\bibinfo {author} {\bibfnamefont {K.}~\bibnamefont
  {{Balakier}}}, \bibinfo {author} {\bibfnamefont {H.}~\bibnamefont {{Shams}}},
  \bibinfo {author} {\bibfnamefont {M.~J.}\ \bibnamefont {{Fice}}}, \bibinfo
  {author} {\bibfnamefont {L.}~\bibnamefont {{Ponnampalam}}}, \bibinfo {author}
  {\bibfnamefont {C.~S.}\ \bibnamefont {{Graham}}}, \bibinfo {author}
  {\bibfnamefont {C.~C.}\ \bibnamefont {{Renaud}}}, \ and\ \bibinfo {author}
  {\bibfnamefont {A.~J.}\ \bibnamefont {{Seeds}}},\ }\bibfield  {title}
  {\enquote {\bibinfo {title} {Optical phase lock loop as high-quality tuneable
  filter for optical frequency comb line selection},}\ }\href {\doibase
  10.1109/JLT.2018.2848961} {\bibfield  {journal} {\bibinfo  {journal} {Journal
  of Lightwave Technology}\ }\textbf {\bibinfo {volume} {36}},\ \bibinfo
  {pages} {4646--4654} (\bibinfo {year} {2018})}\BibitemShut {NoStop}%
\bibitem [{\citenamefont {Kauppinen}\ \emph {et~al.}(1981)\citenamefont
  {Kauppinen}, \citenamefont {Moffatt}, \citenamefont {Mantsch},\ and\
  \citenamefont {Cameron}}]{Kauppinen1981}%
  \BibitemOpen
  \bibfield  {author} {\bibinfo {author} {\bibfnamefont {J.~K.}\ \bibnamefont
  {Kauppinen}}, \bibinfo {author} {\bibfnamefont {D.~J.}\ \bibnamefont
  {Moffatt}}, \bibinfo {author} {\bibfnamefont {H.~H.}\ \bibnamefont
  {Mantsch}}, \ and\ \bibinfo {author} {\bibfnamefont {D.~G.}\ \bibnamefont
  {Cameron}},\ }\bibfield  {title} {\enquote {\bibinfo {title} {{Fourier
  Self-Deconvolution: A Method for Resolving Intrinsically Overlapped
  Bands}},}\ }\href {\doibase 10.1366/0003702814732634} {\bibfield  {journal}
  {\bibinfo  {journal} {Applied Spectroscopy}\ }\textbf {\bibinfo {volume}
  {35}},\ \bibinfo {pages} {271--276} (\bibinfo {year} {1981})}\BibitemShut
  {NoStop}%
\bibitem [{\citenamefont {Nagatsuma}, \citenamefont {Ducournau},\ and\
  \citenamefont {Renaud}(2016)}]{Nagatsuma2016}%
  \BibitemOpen
  \bibfield  {author} {\bibinfo {author} {\bibfnamefont {T.}~\bibnamefont
  {Nagatsuma}}, \bibinfo {author} {\bibfnamefont {G.}~\bibnamefont
  {Ducournau}}, \ and\ \bibinfo {author} {\bibfnamefont {C.~C.}\ \bibnamefont
  {Renaud}},\ }\bibfield  {title} {\enquote {\bibinfo {title} {{Advances in
  terahertz communications accelerated by photonics}},}\ }\href
  {https://doi.org/10.1038/nphoton.2016.65 http://10.0.4.14/nphoton.2016.65}
  {\bibfield  {journal} {\bibinfo  {journal} {Nature Photonics}\ }\textbf
  {\bibinfo {volume} {10}},\ \bibinfo {pages} {371} (\bibinfo {year}
  {2016})}\BibitemShut {NoStop}%
\bibitem [{\citenamefont {Ishibashi}\ \emph {et~al.}(1997)\citenamefont
  {Ishibashi}, \citenamefont {Shimizu}, \citenamefont {Kodama}, \citenamefont
  {Ito}, \citenamefont {Nagatsuma}, \citenamefont {Furuta}, \citenamefont
  {Photodiodes}, \citenamefont {Ishibashi}, \citenamefont {Shimizu},
  \citenamefont {Kodama}, \citenamefont {Ito}, \citenamefont {Nagatsuma},\ and\
  \citenamefont {Furuta}}]{FirstUTCPD}%
  \BibitemOpen
  \bibfield  {author} {\bibinfo {author} {\bibfnamefont {T.}~\bibnamefont
  {Ishibashi}}, \bibinfo {author} {\bibfnamefont {N.}~\bibnamefont {Shimizu}},
  \bibinfo {author} {\bibfnamefont {S.}~\bibnamefont {Kodama}}, \bibinfo
  {author} {\bibfnamefont {H.}~\bibnamefont {Ito}}, \bibinfo {author}
  {\bibfnamefont {T.}~\bibnamefont {Nagatsuma}}, \bibinfo {author}
  {\bibfnamefont {T.}~\bibnamefont {Furuta}}, \bibinfo {author} {\bibfnamefont
  {U.-t.-c.}\ \bibnamefont {Photodiodes}}, \bibinfo {author} {\bibfnamefont
  {T.}~\bibnamefont {Ishibashi}}, \bibinfo {author} {\bibfnamefont
  {N.}~\bibnamefont {Shimizu}}, \bibinfo {author} {\bibfnamefont
  {S.}~\bibnamefont {Kodama}}, \bibinfo {author} {\bibfnamefont
  {H.}~\bibnamefont {Ito}}, \bibinfo {author} {\bibfnamefont {T.}~\bibnamefont
  {Nagatsuma}}, \ and\ \bibinfo {author} {\bibfnamefont {T.}~\bibnamefont
  {Furuta}},\ }\bibfield  {title} {\enquote {\bibinfo {title}
  {{Uni-Traveling-Carrier Photodiodes}},}\ }\href@noop {} {\bibfield  {journal}
  {\bibinfo  {journal} {Ultrafast Electronics and Optoelectronics}\ }\textbf
  {\bibinfo {volume} {13}},\ \bibinfo {pages} {83--87} (\bibinfo {year}
  {1997})}\BibitemShut {NoStop}%
\bibitem [{\citenamefont {Rouvalis}\ \emph {et~al.}(2011)\citenamefont
  {Rouvalis}, \citenamefont {Fice}, \citenamefont {Renaud},\ and\ \citenamefont
  {Seeds}}]{Rouvalis2011}%
  \BibitemOpen
  \bibfield  {author} {\bibinfo {author} {\bibfnamefont {E.}~\bibnamefont
  {Rouvalis}}, \bibinfo {author} {\bibfnamefont {M.~J.}\ \bibnamefont {Fice}},
  \bibinfo {author} {\bibfnamefont {C.~C.}\ \bibnamefont {Renaud}}, \ and\
  \bibinfo {author} {\bibfnamefont {A.~J.}\ \bibnamefont {Seeds}},\ }\bibfield
  {title} {\enquote {\bibinfo {title} {{Optoelectronic detection of
  millimetre-wave signals with travelling-wave uni-travelling carrier
  photodiodes}},}\ }\href
  {http://www.opticsexpress.org/abstract.cfm?URI=oe-19-3-2079} {\bibfield
  {journal} {\bibinfo  {journal} {Opt. Express}\ }\textbf {\bibinfo {volume}
  {19}},\ \bibinfo {pages} {2079--2084} (\bibinfo {year} {2011})}\BibitemShut
  {NoStop}%
\bibitem [{\citenamefont {Latzel}\ \emph {et~al.}(2017)\citenamefont {Latzel},
  \citenamefont {Pavanello}, \citenamefont {Billet}, \citenamefont {Bretin},
  \citenamefont {Beck}, \citenamefont {Vanwolleghem}, \citenamefont {Coinon},
  \citenamefont {Wallart}, \citenamefont {Peytavit}, \citenamefont {Ducournau},
  \citenamefont {Zaknoune},\ and\ \citenamefont {Lampin}}]{LampinCavityUTC}%
  \BibitemOpen
  \bibfield  {author} {\bibinfo {author} {\bibfnamefont {P.}~\bibnamefont
  {Latzel}}, \bibinfo {author} {\bibfnamefont {F.}~\bibnamefont {Pavanello}},
  \bibinfo {author} {\bibfnamefont {M.}~\bibnamefont {Billet}}, \bibinfo
  {author} {\bibfnamefont {S.}~\bibnamefont {Bretin}}, \bibinfo {author}
  {\bibfnamefont {A.}~\bibnamefont {Beck}}, \bibinfo {author} {\bibfnamefont
  {M.}~\bibnamefont {Vanwolleghem}}, \bibinfo {author} {\bibfnamefont
  {C.}~\bibnamefont {Coinon}}, \bibinfo {author} {\bibfnamefont
  {X.}~\bibnamefont {Wallart}}, \bibinfo {author} {\bibfnamefont
  {E.}~\bibnamefont {Peytavit}}, \bibinfo {author} {\bibfnamefont
  {G.}~\bibnamefont {Ducournau}}, \bibinfo {author} {\bibfnamefont
  {M.}~\bibnamefont {Zaknoune}}, \ and\ \bibinfo {author} {\bibfnamefont
  {J.}~\bibnamefont {Lampin}},\ }\bibfield  {title} {\enquote {\bibinfo {title}
  {{Generation of mW Level in the 300-GHz Band Using Resonant-Cavity-Enhanced
  Unitraveling Carrier Photodiodes}},}\ }\href {\doibase
  10.1109/TTHZ.2017.2756059} {\bibfield  {journal} {\bibinfo  {journal} {IEEE
  Transactions on Terahertz Science and Technology}\ }\textbf {\bibinfo
  {volume} {7}},\ \bibinfo {pages} {800--807} (\bibinfo {year}
  {2017})}\BibitemShut {NoStop}%
\bibitem [{\citenamefont {Renaud}\ \emph {et~al.}(2006)\citenamefont {Renaud},
  \citenamefont {Robertson}, \citenamefont {Rogers}, \citenamefont {Firth},
  \citenamefont {Cannard}, \citenamefont {Moore},\ and\ \citenamefont
  {Seeds}}]{renaud2006high}%
  \BibitemOpen
  \bibfield  {author} {\bibinfo {author} {\bibfnamefont {C.~C.}\ \bibnamefont
  {Renaud}}, \bibinfo {author} {\bibfnamefont {M.}~\bibnamefont {Robertson}},
  \bibinfo {author} {\bibfnamefont {D.}~\bibnamefont {Rogers}}, \bibinfo
  {author} {\bibfnamefont {R.}~\bibnamefont {Firth}}, \bibinfo {author}
  {\bibfnamefont {P.~J.}\ \bibnamefont {Cannard}}, \bibinfo {author}
  {\bibfnamefont {R.}~\bibnamefont {Moore}}, \ and\ \bibinfo {author}
  {\bibfnamefont {A.~J.}\ \bibnamefont {Seeds}},\ }\bibfield  {title} {\enquote
  {\bibinfo {title} {{A high responsivity, broadband waveguide uni-travelling
  carrier photodiode}},}\ }in\ \href@noop {} {\emph {\bibinfo {booktitle}
  {Millimeter-Wave and Terahertz Photonics}}},\ Vol.\ \bibinfo {volume} {6194}\
  (\bibinfo {organization} {International Society for Optics and Photonics},\
  \bibinfo {year} {2006})\ p.\ \bibinfo {pages} {61940C}\BibitemShut {NoStop}%
\bibitem [{\citenamefont {Rouvalis}\ \emph {et~al.}(2010)\citenamefont
  {Rouvalis}, \citenamefont {Renaud}, \citenamefont {Moodie}, \citenamefont
  {Robertson},\ and\ \citenamefont {Seeds}}]{Rouvalis2010}%
  \BibitemOpen
  \bibfield  {author} {\bibinfo {author} {\bibfnamefont {E.}~\bibnamefont
  {Rouvalis}}, \bibinfo {author} {\bibfnamefont {C.~C.}\ \bibnamefont
  {Renaud}}, \bibinfo {author} {\bibfnamefont {D.~G.}\ \bibnamefont {Moodie}},
  \bibinfo {author} {\bibfnamefont {M.~J.}\ \bibnamefont {Robertson}}, \ and\
  \bibinfo {author} {\bibfnamefont {A.~J.}\ \bibnamefont {Seeds}},\ }\bibfield
  {title} {\enquote {\bibinfo {title} {{Traveling-wave Uni-Traveling Carrier
  Photodiodes for continuous wave THz generation}},}\ }\href {\doibase
  10.1364/OE.18.011105} {\bibfield  {journal} {\bibinfo  {journal} {Opt.
  Express}\ }\textbf {\bibinfo {volume} {18}},\ \bibinfo {pages} {11105--11110}
  (\bibinfo {year} {2010})}\BibitemShut {NoStop}%
\bibitem [{\citenamefont {Del'Haye}\ \emph {et~al.}(2007)\citenamefont
  {Del'Haye}, \citenamefont {Schliesser}, \citenamefont {Arcizet},
  \citenamefont {Wilken}, \citenamefont {Holzwarth},\ and\ \citenamefont
  {Kippenberg}}]{del2007optical}%
  \BibitemOpen
  \bibfield  {author} {\bibinfo {author} {\bibfnamefont {P.}~\bibnamefont
  {Del'Haye}}, \bibinfo {author} {\bibfnamefont {A.}~\bibnamefont
  {Schliesser}}, \bibinfo {author} {\bibfnamefont {O.}~\bibnamefont {Arcizet}},
  \bibinfo {author} {\bibfnamefont {T.}~\bibnamefont {Wilken}}, \bibinfo
  {author} {\bibfnamefont {R.}~\bibnamefont {Holzwarth}}, \ and\ \bibinfo
  {author} {\bibfnamefont {T.~J.}\ \bibnamefont {Kippenberg}},\ }\bibfield
  {title} {\enquote {\bibinfo {title} {{Optical frequency comb generation from
  a monolithic microresonator}},}\ }\href@noop {} {\bibfield  {journal}
  {\bibinfo  {journal} {Nature}\ }\textbf {\bibinfo {volume} {450}},\ \bibinfo
  {pages} {1214} (\bibinfo {year} {2007})}\BibitemShut {NoStop}%
\bibitem [{\citenamefont {Kues}\ \emph {et~al.}(2019)\citenamefont {Kues},
  \citenamefont {Reimer}, \citenamefont {Lukens}, \citenamefont {Munro},
  \citenamefont {Weiner}, \citenamefont {Moss},\ and\ \citenamefont
  {Morandotti}}]{Kues2019}%
  \BibitemOpen
  \bibfield  {author} {\bibinfo {author} {\bibfnamefont {M.}~\bibnamefont
  {Kues}}, \bibinfo {author} {\bibfnamefont {C.}~\bibnamefont {Reimer}},
  \bibinfo {author} {\bibfnamefont {J.~M.}\ \bibnamefont {Lukens}}, \bibinfo
  {author} {\bibfnamefont {W.~J.}\ \bibnamefont {Munro}}, \bibinfo {author}
  {\bibfnamefont {A.~M.}\ \bibnamefont {Weiner}}, \bibinfo {author}
  {\bibfnamefont {D.~J.}\ \bibnamefont {Moss}}, \ and\ \bibinfo {author}
  {\bibfnamefont {R.}~\bibnamefont {Morandotti}},\ }\bibfield  {title}
  {\enquote {\bibinfo {title} {{Quantum optical microcombs}},}\ }\href
  {\doibase 10.1038/s41566-019-0363-0} {\bibfield  {journal} {\bibinfo
  {journal} {Nature Photonics}\ }\textbf {\bibinfo {volume} {13}},\ \bibinfo
  {pages} {170--179} (\bibinfo {year} {2019})}\BibitemShut {NoStop}%
\bibitem [{\citenamefont {Sanjoh}\ \emph {et~al.}(1997)\citenamefont {Sanjoh},
  \citenamefont {Yasaka}, \citenamefont {Sakai}, \citenamefont {Sato},
  \citenamefont {Ishii},\ and\ \citenamefont {Yoshikuni}}]{MLLComb}%
  \BibitemOpen
  \bibfield  {author} {\bibinfo {author} {\bibfnamefont {H.}~\bibnamefont
  {Sanjoh}}, \bibinfo {author} {\bibfnamefont {H.}~\bibnamefont {Yasaka}},
  \bibinfo {author} {\bibfnamefont {Y.}~\bibnamefont {Sakai}}, \bibinfo
  {author} {\bibfnamefont {K.}~\bibnamefont {Sato}}, \bibinfo {author}
  {\bibfnamefont {H.}~\bibnamefont {Ishii}}, \ and\ \bibinfo {author}
  {\bibfnamefont {Y.}~\bibnamefont {Yoshikuni}},\ }\bibfield  {title} {\enquote
  {\bibinfo {title} {{Multiwavelength light source with precise frequency
  spacing using a mode-locked semiconductor laser and an arrayed waveguide
  grating filter}},}\ }\href {\doibase 10.1109/68.585001} {\bibfield  {journal}
  {\bibinfo  {journal} {IEEE Photonics Technology Letters}\ }\textbf {\bibinfo
  {volume} {9}},\ \bibinfo {pages} {818--820} (\bibinfo {year}
  {1997})}\BibitemShut {NoStop}%
\bibitem [{\citenamefont {Ho}\ and\ \citenamefont
  {Kahn}(1993)}]{ModulatorComb}%
  \BibitemOpen
  \bibfield  {author} {\bibinfo {author} {\bibfnamefont {K.~.}\ \bibnamefont
  {Ho}}\ and\ \bibinfo {author} {\bibfnamefont {J.~M.}\ \bibnamefont {Kahn}},\
  }\bibfield  {title} {\enquote {\bibinfo {title} {{Optical frequency comb
  generator using phase modulation in amplified circulating loop}},}\ }\href
  {\doibase 10.1109/68.219723} {\bibfield  {journal} {\bibinfo  {journal} {IEEE
  Photonics Technology Letters}\ }\textbf {\bibinfo {volume} {5}},\ \bibinfo
  {pages} {721--725} (\bibinfo {year} {1993})}\BibitemShut {NoStop}%
\bibitem [{\citenamefont {Bennett}\ \emph {et~al.}(1999)\citenamefont
  {Bennett}, \citenamefont {Cai}, \citenamefont {Burr}, \citenamefont {Gough},\
  and\ \citenamefont {Seeds}}]{Bennett1999}%
  \BibitemOpen
  \bibfield  {author} {\bibinfo {author} {\bibfnamefont {S.}~\bibnamefont
  {Bennett}}, \bibinfo {author} {\bibfnamefont {B.}~\bibnamefont {Cai}},
  \bibinfo {author} {\bibfnamefont {E.}~\bibnamefont {Burr}}, \bibinfo {author}
  {\bibfnamefont {O.}~\bibnamefont {Gough}}, \ and\ \bibinfo {author}
  {\bibfnamefont {A.~J.}\ \bibnamefont {Seeds}},\ }\bibfield  {title} {\enquote
  {\bibinfo {title} {{1.8-THz bandwidth, zero-frequency error, tunable optical
  comb generator for DWDM applications}},}\ }\href {\doibase 10.1109/68.759395}
  {\bibfield  {journal} {\bibinfo  {journal} {IEEE Photonics Technology
  Letters}\ }\textbf {\bibinfo {volume} {11}},\ \bibinfo {pages} {551--553}
  (\bibinfo {year} {1999})}\BibitemShut {NoStop}%
\bibitem [{\citenamefont {Kourogi}, \citenamefont {Enami},\ and\ \citenamefont
  {Ohtsu}(1994)}]{monolithComb}%
  \BibitemOpen
  \bibfield  {author} {\bibinfo {author} {\bibfnamefont {M.}~\bibnamefont
  {Kourogi}}, \bibinfo {author} {\bibfnamefont {T.}~\bibnamefont {Enami}}, \
  and\ \bibinfo {author} {\bibfnamefont {M.}~\bibnamefont {Ohtsu}},\ }\bibfield
   {title} {\enquote {\bibinfo {title} {{A monolithic optical frequency comb
  generator}},}\ }\href {\doibase 10.1109/68.275432} {\bibfield  {journal}
  {\bibinfo  {journal} {IEEE Photonics Technology Letters}\ }\textbf {\bibinfo
  {volume} {6}},\ \bibinfo {pages} {214--217} (\bibinfo {year}
  {1994})}\BibitemShut {NoStop}%
\bibitem [{\citenamefont {Ponnampalam}\ \emph {et~al.}(2018)\citenamefont
  {Ponnampalam}, \citenamefont {Fice}, \citenamefont {Shams}, \citenamefont
  {Renaud},\ and\ \citenamefont {Seeds}}]{PonnampalamComb}%
  \BibitemOpen
  \bibfield  {author} {\bibinfo {author} {\bibfnamefont {L.}~\bibnamefont
  {Ponnampalam}}, \bibinfo {author} {\bibfnamefont {M.}~\bibnamefont {Fice}},
  \bibinfo {author} {\bibfnamefont {H.}~\bibnamefont {Shams}}, \bibinfo
  {author} {\bibfnamefont {C.}~\bibnamefont {Renaud}}, \ and\ \bibinfo {author}
  {\bibfnamefont {A.}~\bibnamefont {Seeds}},\ }\bibfield  {title} {\enquote
  {\bibinfo {title} {{Optical comb for generation of a continuously tunable
  coherent THz signal from 122.5{\&}{\#}x2009;{\&}{\#}x2009;GHz to
  {\textgreater}2.7{\&}{\#}x00A0;THz}},}\ }\href {\doibase
  10.1364/OL.43.002507} {\bibfield  {journal} {\bibinfo  {journal} {Opt.
  Lett.}\ }\textbf {\bibinfo {volume} {43}},\ \bibinfo {pages} {2507--2510}
  (\bibinfo {year} {2018})}\BibitemShut {NoStop}%
\bibitem [{\citenamefont {Takita}\ \emph {et~al.}(2004)\citenamefont {Takita},
  \citenamefont {Futami}, \citenamefont {Doi},\ and\ \citenamefont
  {Watanabe}}]{Takita:04}%
  \BibitemOpen
  \bibfield  {author} {\bibinfo {author} {\bibfnamefont {Y.}~\bibnamefont
  {Takita}}, \bibinfo {author} {\bibfnamefont {F.}~\bibnamefont {Futami}},
  \bibinfo {author} {\bibfnamefont {M.}~\bibnamefont {Doi}}, \ and\ \bibinfo
  {author} {\bibfnamefont {S.}~\bibnamefont {Watanabe}},\ }\bibfield  {title}
  {\enquote {\bibinfo {title} {{Highly stable ultra-short pulse generation by
  filtering out flat optical frequency components}},}\ }in\ \href
  {http://www.osapublishing.org/abstract.cfm?URI=CLEO-2004-CTuN1} {\emph
  {\bibinfo {booktitle} {Conference on Lasers and Electro-Optics/International
  Quantum Electronics Conference and Photonic Applications Systems
  Technologies}}}\ (\bibinfo  {publisher} {Optical Society of America},\
  \bibinfo {year} {2004})\ p.\ \bibinfo {pages} {CTuN1}\BibitemShut {NoStop}%
\bibitem [{\citenamefont {Sakamoto}, \citenamefont {Kawanishi},\ and\
  \citenamefont {Izutsu}(2007)}]{SakamotoComb}%
  \BibitemOpen
  \bibfield  {author} {\bibinfo {author} {\bibfnamefont {T.}~\bibnamefont
  {Sakamoto}}, \bibinfo {author} {\bibfnamefont {T.}~\bibnamefont {Kawanishi}},
  \ and\ \bibinfo {author} {\bibfnamefont {M.}~\bibnamefont {Izutsu}},\
  }\bibfield  {title} {\enquote {\bibinfo {title} {{Widely wavelength-tunable
  ultra-flat frequency comb generation using conventional dual-drive
  Mach-Zehnder modulator}},}\ }\href {\doibase 10.1049/el:20071267} {\bibfield
  {journal} {\bibinfo  {journal} {Electronics Letters}\ }\textbf {\bibinfo
  {volume} {43}},\ \bibinfo {pages} {1039--1040} (\bibinfo {year}
  {2007})}\BibitemShut {NoStop}%
\bibitem [{\citenamefont {Magari{\~{n}}o}\ \emph {et~al.}(1976)\citenamefont
  {Magari{\~{n}}o}, \citenamefont {Tuchendler}, \citenamefont {D'Haenens},\
  and\ \citenamefont {Linz}}]{PhysRevB.13.2805}%
  \BibitemOpen
  \bibfield  {author} {\bibinfo {author} {\bibfnamefont {J.}~\bibnamefont
  {Magari{\~{n}}o}}, \bibinfo {author} {\bibfnamefont {J.}~\bibnamefont
  {Tuchendler}}, \bibinfo {author} {\bibfnamefont {J.~P.}\ \bibnamefont
  {D'Haenens}}, \ and\ \bibinfo {author} {\bibfnamefont {A.}~\bibnamefont
  {Linz}},\ }\bibfield  {title} {\enquote {\bibinfo {title} {{Submillimeter
  resonance spectroscopy of $\backslash$mathrm{\{}Ho{\^{}}{\{}3+{\}}{\}} in
  lithium yttrium fluoride}},}\ }\href {\doibase 10.1103/PhysRevB.13.2805}
  {\bibfield  {journal} {\bibinfo  {journal} {Phys. Rev. B}\ }\textbf {\bibinfo
  {volume} {13}},\ \bibinfo {pages} {2805--2808} (\bibinfo {year}
  {1976})}\BibitemShut {NoStop}%
\bibitem [{\citenamefont {Santo}\ \emph {et~al.}(2006)\citenamefont {Santo},
  \citenamefont {Librantz}, \citenamefont {Gomes}, \citenamefont {Pizani},
  \citenamefont {Ranieri}, \citenamefont {Vieira},\ and\ \citenamefont
  {Baldochi}}]{Santo2006b}%
  \BibitemOpen
  \bibfield  {author} {\bibinfo {author} {\bibfnamefont {A.}~\bibnamefont
  {Santo}}, \bibinfo {author} {\bibfnamefont {A.}~\bibnamefont {Librantz}},
  \bibinfo {author} {\bibfnamefont {L.}~\bibnamefont {Gomes}}, \bibinfo
  {author} {\bibfnamefont {P.}~\bibnamefont {Pizani}}, \bibinfo {author}
  {\bibfnamefont {I.}~\bibnamefont {Ranieri}}, \bibinfo {author} {\bibfnamefont
  {N.}~\bibnamefont {Vieira}}, \ and\ \bibinfo {author} {\bibfnamefont
  {S.}~\bibnamefont {Baldochi}},\ }\bibfield  {title} {\enquote {\bibinfo
  {title} {{Growth and characterization of LiYF4:Nd single crystal fibres for
  optical applications}},}\ }\href {\doibase 10.1016/j.jcrysgro.2006.03.054}
  {\bibfield  {journal} {\bibinfo  {journal} {Journal of Crystal Growth}\
  }\textbf {\bibinfo {volume} {292}},\ \bibinfo {pages} {149--154} (\bibinfo
  {year} {2006})}\BibitemShut {NoStop}%
\bibitem [{\citenamefont {Reich}\ \emph {et~al.}(1990)\citenamefont {Reich},
  \citenamefont {Ellman}, \citenamefont {Yang}, \citenamefont {Rosenbaum},
  \citenamefont {Aeppli},\ and\ \citenamefont {Belanger}}]{Reich1990}%
  \BibitemOpen
  \bibfield  {author} {\bibinfo {author} {\bibfnamefont {D.~H.}\ \bibnamefont
  {Reich}}, \bibinfo {author} {\bibfnamefont {B.}~\bibnamefont {Ellman}},
  \bibinfo {author} {\bibfnamefont {J.}~\bibnamefont {Yang}}, \bibinfo {author}
  {\bibfnamefont {T.~F.}\ \bibnamefont {Rosenbaum}}, \bibinfo {author}
  {\bibfnamefont {G.}~\bibnamefont {Aeppli}}, \ and\ \bibinfo {author}
  {\bibfnamefont {D.~P.}\ \bibnamefont {Belanger}},\ }\bibfield  {title}
  {\enquote {\bibinfo {title} {{Dipolar magnets and glasses:
  Neutron-scattering, dynamical, and calorimetric studies of randomly
  distributed Ising spins}},}\ }\href {\doibase 10.1103/PhysRevB.42.4631}
  {\bibfield  {journal} {\bibinfo  {journal} {Phys. Rev. B}\ }\textbf {\bibinfo
  {volume} {42}},\ \bibinfo {pages} {4631--4644} (\bibinfo {year}
  {1990})}\BibitemShut {NoStop}%
\bibitem [{\citenamefont {Brooke}(1999)}]{Brooke1999}%
  \BibitemOpen
  \bibfield  {author} {\bibinfo {author} {\bibfnamefont {J.}~\bibnamefont
  {Brooke}},\ }\bibfield  {title} {\enquote {\bibinfo {title} {{Quantum
  Annealing of a Disordered Magnet}},}\ }\href {\doibase
  10.1126/science.284.5415.779} {\bibfield  {journal} {\bibinfo  {journal}
  {Science}\ }\textbf {\bibinfo {volume} {284}},\ \bibinfo {pages} {779--781}
  (\bibinfo {year} {1999})}\BibitemShut {NoStop}%
\bibitem [{\citenamefont {R{\o}nnow}\ \emph {et~al.}(2007)\citenamefont
  {R{\o}nnow}, \citenamefont {Jensen}, \citenamefont {Parthasarathy},
  \citenamefont {Aeppli}, \citenamefont {Rosenbaum}, \citenamefont {McMorrow},\
  and\ \citenamefont {Kraemer}}]{Ronnow2007}%
  \BibitemOpen
  \bibfield  {author} {\bibinfo {author} {\bibfnamefont {H.~M.}\ \bibnamefont
  {R{\o}nnow}}, \bibinfo {author} {\bibfnamefont {J.}~\bibnamefont {Jensen}},
  \bibinfo {author} {\bibfnamefont {R.}~\bibnamefont {Parthasarathy}}, \bibinfo
  {author} {\bibfnamefont {G.}~\bibnamefont {Aeppli}}, \bibinfo {author}
  {\bibfnamefont {T.~F.}\ \bibnamefont {Rosenbaum}}, \bibinfo {author}
  {\bibfnamefont {D.~F.}\ \bibnamefont {McMorrow}}, \ and\ \bibinfo {author}
  {\bibfnamefont {C.}~\bibnamefont {Kraemer}},\ }\bibfield  {title} {\enquote
  {\bibinfo {title} {{Magnetic excitations near the quantum phase transition in
  the Ising ferromagnet LiHo F4}},}\ }\href {\doibase
  10.1103/PhysRevB.75.054426} {\bibfield  {journal} {\bibinfo  {journal} {Phys.
  Rev. B - Condens. Matter Mater. Phys.}\ }\textbf {\bibinfo {volume} {75}},\
  \bibinfo {pages} {1--8} (\bibinfo {year} {2007})}\BibitemShut {NoStop}%
\bibitem [{\citenamefont {Brooke}, \citenamefont {Rosenbaum},\ and\
  \citenamefont {Aeppli}(2001)}]{Brooke2001}%
  \BibitemOpen
  \bibfield  {author} {\bibinfo {author} {\bibfnamefont {J.}~\bibnamefont
  {Brooke}}, \bibinfo {author} {\bibfnamefont {T.~F.}\ \bibnamefont
  {Rosenbaum}}, \ and\ \bibinfo {author} {\bibfnamefont {G.}~\bibnamefont
  {Aeppli}},\ }\bibfield  {title} {\enquote {\bibinfo {title} {{Tunable quantum
  tunnelling of magnetic domain walls.}}}\ }\href {\doibase 10.1038/35098037}
  {\bibfield  {journal} {\bibinfo  {journal} {Nature}\ }\textbf {\bibinfo
  {volume} {413}},\ \bibinfo {pages} {610--3} (\bibinfo {year}
  {2001})}\BibitemShut {NoStop}%
\bibitem [{\citenamefont {Ghosh}\ \emph {et~al.}(2002)\citenamefont {Ghosh},
  \citenamefont {Parthasarathy}, \citenamefont {Rosenbaum},\ and\ \citenamefont
  {Aeppli}}]{Ghosh2002}%
  \BibitemOpen
  \bibfield  {author} {\bibinfo {author} {\bibfnamefont {S.}~\bibnamefont
  {Ghosh}}, \bibinfo {author} {\bibfnamefont {R.}~\bibnamefont
  {Parthasarathy}}, \bibinfo {author} {\bibfnamefont {T.~F.}\ \bibnamefont
  {Rosenbaum}}, \ and\ \bibinfo {author} {\bibfnamefont {G.}~\bibnamefont
  {Aeppli}},\ }\bibfield  {title} {\enquote {\bibinfo {title} {{Coherent spin
  oscillations in a disordered magnet.}}}\ }\href {\doibase
  10.1126/science.1070731} {\bibfield  {journal} {\bibinfo  {journal}
  {Science}\ }\textbf {\bibinfo {volume} {296}},\ \bibinfo {pages} {2195--2198}
  (\bibinfo {year} {2002})},\ \Eprint {http://arxiv.org/abs/0305541}
  {arXiv:0305541 [cond-mat]} \BibitemShut {NoStop}%
\bibitem [{\citenamefont {Ghosh}\ \emph {et~al.}(2003)\citenamefont {Ghosh},
  \citenamefont {Rosenbaum}, \citenamefont {Aeppli},\ and\ \citenamefont
  {Coppersmith}}]{Ghosh2003a}%
  \BibitemOpen
  \bibfield  {author} {\bibinfo {author} {\bibfnamefont {S.}~\bibnamefont
  {Ghosh}}, \bibinfo {author} {\bibfnamefont {T.~F.}\ \bibnamefont
  {Rosenbaum}}, \bibinfo {author} {\bibfnamefont {G.}~\bibnamefont {Aeppli}}, \
  and\ \bibinfo {author} {\bibfnamefont {S.~N.}\ \bibnamefont {Coppersmith}},\
  }\bibfield  {title} {\enquote {\bibinfo {title} {{Entangled quantum state of
  magnetic dipoles}},}\ }\href {\doibase 10.1038/nature01888} {\bibfield
  {journal} {\bibinfo  {journal} {Nature}\ }\textbf {\bibinfo {volume} {425}},\
  \bibinfo {pages} {48--51} (\bibinfo {year} {2003})}\BibitemShut {NoStop}%
\bibitem [{\citenamefont {Bitko}, \citenamefont {Rosenbaum},\ and\
  \citenamefont {Aeppli}(1996)}]{Bitko1996}%
  \BibitemOpen
  \bibfield  {author} {\bibinfo {author} {\bibfnamefont {D.}~\bibnamefont
  {Bitko}}, \bibinfo {author} {\bibfnamefont {T.~F.}\ \bibnamefont
  {Rosenbaum}}, \ and\ \bibinfo {author} {\bibfnamefont {G.}~\bibnamefont
  {Aeppli}},\ }\bibfield  {title} {\enquote {\bibinfo {title} {{Quantum
  Critical Behavior for a Model Magnet}},}\ }\href {\doibase
  10.1103/PhysRevLett.77.940} {\bibfield  {journal} {\bibinfo  {journal}
  {Physical Review Letters}\ }\textbf {\bibinfo {volume} {77}},\ \bibinfo
  {pages} {940--943} (\bibinfo {year} {1996})}\BibitemShut {NoStop}%
\bibitem [{\citenamefont {R{\o}nnow}\ \emph {et~al.}(2005)\citenamefont
  {R{\o}nnow}, \citenamefont {Parthasarathy}, \citenamefont {Jensen},
  \citenamefont {Aeppli}, \citenamefont {Rosenbaum},\ and\ \citenamefont
  {McMorrow}}]{Ronnow2005}%
  \BibitemOpen
  \bibfield  {author} {\bibinfo {author} {\bibfnamefont {H.~M.}\ \bibnamefont
  {R{\o}nnow}}, \bibinfo {author} {\bibfnamefont {R.}~\bibnamefont
  {Parthasarathy}}, \bibinfo {author} {\bibfnamefont {J.}~\bibnamefont
  {Jensen}}, \bibinfo {author} {\bibfnamefont {G.}~\bibnamefont {Aeppli}},
  \bibinfo {author} {\bibfnamefont {T.~F.}\ \bibnamefont {Rosenbaum}}, \ and\
  \bibinfo {author} {\bibfnamefont {D.~F.}\ \bibnamefont {McMorrow}},\
  }\bibfield  {title} {\enquote {\bibinfo {title} {{Quantum Phase Transition of
  a Magnet in a Spin Bath}},}\ }\href {\doibase 10.1126/science.1108317}
  {\bibfield  {journal} {\bibinfo  {journal} {Science}\ }\textbf {\bibinfo
  {volume} {308}},\ \bibinfo {pages} {389--392} (\bibinfo {year}
  {2005})}\BibitemShut {NoStop}%
\bibitem [{\citenamefont {Bertaina}\ \emph {et~al.}(2007)\citenamefont
  {Bertaina}, \citenamefont {Gambarelli}, \citenamefont {Tkachuk},
  \citenamefont {Kurkin}, \citenamefont {Malkin}, \citenamefont {Stepanov},\
  and\ \citenamefont {Barbara}}]{Bertaina2007}%
  \BibitemOpen
  \bibfield  {author} {\bibinfo {author} {\bibfnamefont {S.}~\bibnamefont
  {Bertaina}}, \bibinfo {author} {\bibfnamefont {S.}~\bibnamefont
  {Gambarelli}}, \bibinfo {author} {\bibfnamefont {A.}~\bibnamefont {Tkachuk}},
  \bibinfo {author} {\bibfnamefont {I.~N.}\ \bibnamefont {Kurkin}}, \bibinfo
  {author} {\bibfnamefont {B.}~\bibnamefont {Malkin}}, \bibinfo {author}
  {\bibfnamefont {A.}~\bibnamefont {Stepanov}}, \ and\ \bibinfo {author}
  {\bibfnamefont {B.}~\bibnamefont {Barbara}},\ }\bibfield  {title} {\enquote
  {\bibinfo {title} {{Rare-earth solid-state qubits}},}\ }\href
  {https://doi.org/10.1038/nnano.2006.174 http://10.0.4.14/nnano.2006.174}
  {\bibfield  {journal} {\bibinfo  {journal} {Nature Nanotechnology}\ }\textbf
  {\bibinfo {volume} {2}},\ \bibinfo {pages} {39} (\bibinfo {year}
  {2007})}\BibitemShut {NoStop}%
\bibitem [{\citenamefont {Bussi{\`{e}}res}\ \emph {et~al.}(2014)\citenamefont
  {Bussi{\`{e}}res}, \citenamefont {Clausen}, \citenamefont {Tiranov},
  \citenamefont {Korzh}, \citenamefont {Verma}, \citenamefont {Nam},
  \citenamefont {Marsili}, \citenamefont {Ferrier}, \citenamefont {Goldner},
  \citenamefont {Herrmann}, \citenamefont {Silberhorn}, \citenamefont {Sohler},
  \citenamefont {Afzelius},\ and\ \citenamefont {Gisin}}]{Bussieres2014}%
  \BibitemOpen
  \bibfield  {author} {\bibinfo {author} {\bibfnamefont {F.}~\bibnamefont
  {Bussi{\`{e}}res}}, \bibinfo {author} {\bibfnamefont {C.}~\bibnamefont
  {Clausen}}, \bibinfo {author} {\bibfnamefont {A.}~\bibnamefont {Tiranov}},
  \bibinfo {author} {\bibfnamefont {B.}~\bibnamefont {Korzh}}, \bibinfo
  {author} {\bibfnamefont {V.~B.}\ \bibnamefont {Verma}}, \bibinfo {author}
  {\bibfnamefont {S.~W.}\ \bibnamefont {Nam}}, \bibinfo {author} {\bibfnamefont
  {F.}~\bibnamefont {Marsili}}, \bibinfo {author} {\bibfnamefont
  {A.}~\bibnamefont {Ferrier}}, \bibinfo {author} {\bibfnamefont
  {P.}~\bibnamefont {Goldner}}, \bibinfo {author} {\bibfnamefont
  {H.}~\bibnamefont {Herrmann}}, \bibinfo {author} {\bibfnamefont
  {C.}~\bibnamefont {Silberhorn}}, \bibinfo {author} {\bibfnamefont
  {W.}~\bibnamefont {Sohler}}, \bibinfo {author} {\bibfnamefont
  {M.}~\bibnamefont {Afzelius}}, \ and\ \bibinfo {author} {\bibfnamefont
  {N.}~\bibnamefont {Gisin}},\ }\bibfield  {title} {\enquote {\bibinfo {title}
  {{Quantum teleportation from a telecom-wavelength photon to a solid-state
  quantum memory}},}\ }\href {https://doi.org/10.1038/nphoton.2014.215
  http://10.0.4.14/nphoton.2014.215
  https://www.nature.com/articles/nphoton.2014.215{\#}supplementary-information}
  {\bibfield  {journal} {\bibinfo  {journal} {Nature Photonics}\ }\textbf
  {\bibinfo {volume} {8}},\ \bibinfo {pages} {775} (\bibinfo {year}
  {2014})}\BibitemShut {NoStop}%
\bibitem [{\citenamefont {Ran{\v{c}}i{\'{c}}}\ \emph
  {et~al.}(2017)\citenamefont {Ran{\v{c}}i{\'{c}}}, \citenamefont {Hedges},
  \citenamefont {Ahlefeldt},\ and\ \citenamefont {Sellars}}]{Rancic2017}%
  \BibitemOpen
  \bibfield  {author} {\bibinfo {author} {\bibfnamefont {M.}~\bibnamefont
  {Ran{\v{c}}i{\'{c}}}}, \bibinfo {author} {\bibfnamefont {M.~P.}\ \bibnamefont
  {Hedges}}, \bibinfo {author} {\bibfnamefont {R.~L.}\ \bibnamefont
  {Ahlefeldt}}, \ and\ \bibinfo {author} {\bibfnamefont {M.~J.}\ \bibnamefont
  {Sellars}},\ }\bibfield  {title} {\enquote {\bibinfo {title} {{Coherence time
  of over a second in a telecom-compatible quantum memory
  storage material}},}\ }\href {https://doi.org/10.1038/nphys4254
  http://10.0.4.14/nphys4254
  https://www.nature.com/articles/nphys4254{\#}supplementary-information}
  {\bibfield  {journal} {\bibinfo  {journal} {Nature Physics}\ }\textbf
  {\bibinfo {volume} {14}},\ \bibinfo {pages} {50} (\bibinfo {year}
  {2017})}\BibitemShut {NoStop}%
\bibitem [{\citenamefont {Karayianis}, \citenamefont {Wortman},\ and\
  \citenamefont {Jenssen}(1976)}]{Karayianis1976}%
  \BibitemOpen
  \bibfield  {author} {\bibinfo {author} {\bibfnamefont {N.}~\bibnamefont
  {Karayianis}}, \bibinfo {author} {\bibfnamefont {D.~E.}\ \bibnamefont
  {Wortman}}, \ and\ \bibinfo {author} {\bibfnamefont {H.~P.}\ \bibnamefont
  {Jenssen}},\ }\bibfield  {title} {\enquote {\bibinfo {title} {{Analysis of
  the optical spectrum of Ho3+ in LiYF4}},}\ }\href {\doibase
  10.1016/0022-3697(76)90004-4} {\bibfield  {journal} {\bibinfo  {journal} {J.
  Phys. Chem. Solids}\ }\textbf {\bibinfo {volume} {37}},\ \bibinfo {pages}
  {675--682} (\bibinfo {year} {1976})}\BibitemShut {NoStop}%
\bibitem [{\citenamefont {Magari{\~{n}}o}\ \emph {et~al.}(1980)\citenamefont
  {Magari{\~{n}}o}, \citenamefont {Tuchendler}, \citenamefont {Beauvillain},\
  and\ \citenamefont {Laursen}}]{PhysRevB.21.18}%
  \BibitemOpen
  \bibfield  {author} {\bibinfo {author} {\bibfnamefont {T.}~\bibnamefont
  {Magari{\~{n}}o}}, \bibinfo {author} {\bibfnamefont {J.}~\bibnamefont
  {Tuchendler}}, \bibinfo {author} {\bibfnamefont {P.}~\bibnamefont
  {Beauvillain}}, \ and\ \bibinfo {author} {\bibfnamefont {I.}~\bibnamefont
  {Laursen}},\ }\bibfield  {title} {\enquote {\bibinfo {title} {{EPR
  experiments in LiTbF{\_}4, LiHoF{\_}4, and LiErF{\_}4 at submillimeter
  frequencies}},}\ }\href {\doibase 10.1103/PhysRevB.21.18} {\bibfield
  {journal} {\bibinfo  {journal} {Phys. Rev. B}\ }\textbf {\bibinfo {volume}
  {21}},\ \bibinfo {pages} {18--28} (\bibinfo {year} {1980})}\BibitemShut
  {NoStop}%
\bibitem [{\citenamefont {Khintchine}(1934)}]{Khintchine1934}%
  \BibitemOpen
  \bibfield  {author} {\bibinfo {author} {\bibfnamefont {A.}~\bibnamefont
  {Khintchine}},\ }\bibfield  {title} {\enquote {\bibinfo {title}
  {{Korrelationstheorie der station{\"{a}}ren stochastischen Prozesse}},}\
  }\href {\doibase 10.1007/BF01449156} {\bibfield  {journal} {\bibinfo
  {journal} {Mathematische Annalen}\ }\textbf {\bibinfo {volume} {109}},\
  \bibinfo {pages} {604--615} (\bibinfo {year} {1934})}\BibitemShut {NoStop}%
\bibitem [{\citenamefont {Koechner}(2013)}]{koechner2013solid}%
  \BibitemOpen
  \bibfield  {author} {\bibinfo {author} {\bibfnamefont {W.}~\bibnamefont
  {Koechner}},\ }\href {https://books.google.co.uk/books?id=NtjqCAAAQBAJ}
  {\emph {\bibinfo {title} {{Solid-State Laser Engineering}}}},\ Springer
  Series in Optical Sciences\ (\bibinfo  {publisher} {Springer Berlin
  Heidelberg},\ \bibinfo {year} {2013})\BibitemShut {NoStop}%
\bibitem [{\citenamefont {Olivero}\ and\ \citenamefont
  {Longbothum}(1977)}]{OLIVERO1977233}%
  \BibitemOpen
  \bibfield  {author} {\bibinfo {author} {\bibfnamefont {J.~J.}\ \bibnamefont
  {Olivero}}\ and\ \bibinfo {author} {\bibfnamefont {R.~L.}\ \bibnamefont
  {Longbothum}},\ }\bibfield  {title} {\enquote {\bibinfo {title} {{Empirical
  fits to the Voigt line width: A brief review}},}\ }\href {\doibase
  https://doi.org/10.1016/0022-4073(77)90161-3} {\bibfield  {journal} {\bibinfo
   {journal} {Journal of Quantitative Spectroscopy and Radiative Transfer}\
  }\textbf {\bibinfo {volume} {17}},\ \bibinfo {pages} {233--236} (\bibinfo
  {year} {1977})}\BibitemShut {NoStop}%
\bibitem [{\citenamefont {He}\ and\ \citenamefont {Zhang}(2013)}]{HE20135245}%
  \BibitemOpen
  \bibfield  {author} {\bibinfo {author} {\bibfnamefont {J.}~\bibnamefont
  {He}}\ and\ \bibinfo {author} {\bibfnamefont {Q.}~\bibnamefont {Zhang}},\
  }\bibfield  {title} {\enquote {\bibinfo {title} {{Discussion on the full
  width at half maximum (FWHM) of the Voigt spectral line}},}\ }\href {\doibase
  https://doi.org/10.1016/j.ijleo.2013.03.173} {\bibfield  {journal} {\bibinfo
  {journal} {Optik}\ }\textbf {\bibinfo {volume} {124}},\ \bibinfo {pages}
  {5245--5247} (\bibinfo {year} {2013})}\BibitemShut {NoStop}%
\bibitem [{\citenamefont {Shams}\ \emph {et~al.}(2016)\citenamefont {Shams},
  \citenamefont {Fice}, \citenamefont {Gonzalez-Guerrero}, \citenamefont
  {Renaud}, \citenamefont {van Dijk},\ and\ \citenamefont {Seeds}}]{Shams:16}%
  \BibitemOpen
  \bibfield  {author} {\bibinfo {author} {\bibfnamefont {H.}~\bibnamefont
  {Shams}}, \bibinfo {author} {\bibfnamefont {M.~J.}\ \bibnamefont {Fice}},
  \bibinfo {author} {\bibfnamefont {L.}~\bibnamefont {Gonzalez-Guerrero}},
  \bibinfo {author} {\bibfnamefont {C.~C.}\ \bibnamefont {Renaud}}, \bibinfo
  {author} {\bibfnamefont {F.}~\bibnamefont {van Dijk}}, \ and\ \bibinfo
  {author} {\bibfnamefont {A.~J.}\ \bibnamefont {Seeds}},\ }\bibfield  {title}
  {\enquote {\bibinfo {title} {{Sub-THz Wireless Over Fiber for Frequency Band
  220–280 GHz}},}\ }\href {\doibase 10.1109/JLT.2016.2558450} {\bibfield
  {journal} {\bibinfo  {journal} {Journal of Lightwave Technology}\ }\textbf
  {\bibinfo {volume} {34}},\ \bibinfo {pages} {4786--4793} (\bibinfo {year}
  {2016})}\BibitemShut {NoStop}%
\bibitem [{\citenamefont {Rouvalis}\ \emph
  {et~al.}(2012{\natexlab{a}})\citenamefont {Rouvalis}, \citenamefont
  {Chtioui}, \citenamefont {Tran}, \citenamefont {Lelarge}, \citenamefont {van
  Dijk}, \citenamefont {Fice}, \citenamefont {Renaud}, \citenamefont
  {Carpintero},\ and\ \citenamefont {Seeds}}]{Rouvalis:12}%
  \BibitemOpen
  \bibfield  {author} {\bibinfo {author} {\bibfnamefont {E.}~\bibnamefont
  {Rouvalis}}, \bibinfo {author} {\bibfnamefont {M.}~\bibnamefont {Chtioui}},
  \bibinfo {author} {\bibfnamefont {M.}~\bibnamefont {Tran}}, \bibinfo {author}
  {\bibfnamefont {F.}~\bibnamefont {Lelarge}}, \bibinfo {author} {\bibfnamefont
  {F.}~\bibnamefont {van Dijk}}, \bibinfo {author} {\bibfnamefont {M.~J.}\
  \bibnamefont {Fice}}, \bibinfo {author} {\bibfnamefont {C.~C.}\ \bibnamefont
  {Renaud}}, \bibinfo {author} {\bibfnamefont {G.}~\bibnamefont {Carpintero}},
  \ and\ \bibinfo {author} {\bibfnamefont {A.~J.}\ \bibnamefont {Seeds}},\
  }\bibfield  {title} {\enquote {\bibinfo {title} {{High-speed photodiodes for
  InP-based photonic integrated circuits}},}\ }\href {\doibase
  10.1364/OE.20.009172} {\bibfield  {journal} {\bibinfo  {journal} {Opt.
  Express}\ }\textbf {\bibinfo {volume} {20}},\ \bibinfo {pages} {9172--9177}
  (\bibinfo {year} {2012}{\natexlab{a}})}\BibitemShut {NoStop}%
\bibitem [{\citenamefont {Ito}\ \emph {et~al.}(2005)\citenamefont {Ito},
  \citenamefont {Nakajima}, \citenamefont {Furuta},\ and\ \citenamefont
  {Ishibashi}}]{Ito_2005}%
  \BibitemOpen
  \bibfield  {author} {\bibinfo {author} {\bibfnamefont {H.}~\bibnamefont
  {Ito}}, \bibinfo {author} {\bibfnamefont {F.}~\bibnamefont {Nakajima}},
  \bibinfo {author} {\bibfnamefont {T.}~\bibnamefont {Furuta}}, \ and\ \bibinfo
  {author} {\bibfnamefont {T.}~\bibnamefont {Ishibashi}},\ }\bibfield  {title}
  {\enquote {\bibinfo {title} {{Continuous {\{}THz{\}}-wave generation using
  antenna-integrated uni-travelling-carrier photodiodes}},}\ }\href {\doibase
  10.1088/0268-1242/20/7/008} {\bibfield  {journal} {\bibinfo  {journal}
  {Semiconductor Science and Technology}\ }\textbf {\bibinfo {volume} {20}},\
  \bibinfo {pages} {S191----S198} (\bibinfo {year} {2005})}\BibitemShut
  {NoStop}%
\bibitem [{\citenamefont {Rouvalis}\ \emph
  {et~al.}(2012{\natexlab{b}})\citenamefont {Rouvalis}, \citenamefont {Renaud},
  \citenamefont {Moodie}, \citenamefont {Robertson},\ and\ \citenamefont
  {Seeds}}]{Rouvalis_2012}%
  \BibitemOpen
  \bibfield  {author} {\bibinfo {author} {\bibfnamefont {E.}~\bibnamefont
  {Rouvalis}}, \bibinfo {author} {\bibfnamefont {C.~C.}\ \bibnamefont
  {Renaud}}, \bibinfo {author} {\bibfnamefont {D.~G.}\ \bibnamefont {Moodie}},
  \bibinfo {author} {\bibfnamefont {M.~J.}\ \bibnamefont {Robertson}}, \ and\
  \bibinfo {author} {\bibfnamefont {A.~J.}\ \bibnamefont {Seeds}},\ }\bibfield
  {title} {\enquote {\bibinfo {title} {{Continuous Wave Terahertz Generation
  From Ultra-Fast InP-Based Photodiodes}},}\ }\href {\doibase
  10.1109/TMTT.2011.2178858} {\bibfield  {journal} {\bibinfo  {journal} {IEEE
  Transactions on Microwave Theory and Techniques}\ }\textbf {\bibinfo {volume}
  {60}},\ \bibinfo {pages} {509--517} (\bibinfo {year}
  {2012}{\natexlab{b}})}\BibitemShut {NoStop}%
\bibitem [{\citenamefont {Hisatake}\ \emph {et~al.}(2014)\citenamefont
  {Hisatake}, \citenamefont {Kim}, \citenamefont {Ajito},\ and\ \citenamefont
  {Nagatsuma}}]{Hisatake:2014}%
  \BibitemOpen
  \bibfield  {author} {\bibinfo {author} {\bibfnamefont {S.}~\bibnamefont
  {Hisatake}}, \bibinfo {author} {\bibfnamefont {J.}~\bibnamefont {Kim}},
  \bibinfo {author} {\bibfnamefont {K.}~\bibnamefont {Ajito}}, \ and\ \bibinfo
  {author} {\bibfnamefont {T.}~\bibnamefont {Nagatsuma}},\ }\bibfield  {title}
  {\enquote {\bibinfo {title} {{Self-Heterodyne Spectrometer Using
  Uni-Traveling-Carrier Photodiodes for Terahertz-Wave Generators and
  Optoelectronic Mixers}},}\ }\href {\doibase 10.1109/JLT.2014.2321004}
  {\bibfield  {journal} {\bibinfo  {journal} {Journal of Lightwave Technology}\
  }\textbf {\bibinfo {volume} {32}},\ \bibinfo {pages} {3683--3689} (\bibinfo
  {year} {2014})}\BibitemShut {NoStop}%
\bibitem [{\citenamefont {Matmon}\ \emph {et~al.}(2015)\citenamefont {Matmon},
  \citenamefont {Lynch}, \citenamefont {Rosenbaum}, \citenamefont {Fisher},
  \citenamefont {Aeppli}, \citenamefont {Kingdom}, \citenamefont {Buildings},
  \citenamefont {Parade}, \citenamefont {Kingdom},\ and\ \citenamefont
  {States}}]{Matmon2015}%
  \BibitemOpen
  \bibfield  {author} {\bibinfo {author} {\bibfnamefont {G.}~\bibnamefont
  {Matmon}}, \bibinfo {author} {\bibfnamefont {S.~A.}\ \bibnamefont {Lynch}},
  \bibinfo {author} {\bibfnamefont {T.~F.}\ \bibnamefont {Rosenbaum}}, \bibinfo
  {author} {\bibfnamefont {A.~J.}\ \bibnamefont {Fisher}}, \bibinfo {author}
  {\bibfnamefont {G.}~\bibnamefont {Aeppli}}, \bibinfo {author} {\bibfnamefont
  {U.}~\bibnamefont {Kingdom}}, \bibinfo {author} {\bibfnamefont
  {Q.}~\bibnamefont {Buildings}}, \bibinfo {author} {\bibfnamefont
  {T.}~\bibnamefont {Parade}}, \bibinfo {author} {\bibfnamefont
  {U.}~\bibnamefont {Kingdom}}, \ and\ \bibinfo {author} {\bibfnamefont
  {U.}~\bibnamefont {States}},\ }\bibfield  {title} {\enquote {\bibinfo {title}
  {{Optical response from THz domain to near-infrared of}},}\ }\href@noop {}
  {\bibfield  {journal} {\bibinfo  {journal} {to be Publ.}\ ,\ \bibinfo {pages}
  {1--27}} (\bibinfo {year} {2015})}\BibitemShut {NoStop}%
\end{thebibliography}

\begin{thebibliography}{1}%
\makeatletter
\providecommand \@ifxundefined [1]{%
 \@ifx{#1\undefined}
}%
\providecommand \@ifnum [1]{%
 \ifnum #1\expandafter \@firstoftwo
 \else \expandafter \@secondoftwo
 \fi
}%
\providecommand \@ifx [1]{%
 \ifx #1\expandafter \@firstoftwo
 \else \expandafter \@secondoftwo
 \fi
}%
\providecommand \natexlab [1]{#1}%
\providecommand \enquote  [1]{``#1''}%
\providecommand \bibnamefont  [1]{#1}%
\providecommand \bibfnamefont [1]{#1}%
\providecommand \citenamefont [1]{#1}%
\providecommand \href@noop [0]{\@secondoftwo}%
\providecommand \href [0]{\begingroup \@sanitize@url \@href}%
\providecommand \@href[1]{\@@startlink{#1}\@@href}%
\providecommand \@@href[1]{\endgroup#1\@@endlink}%
\providecommand \@sanitize@url [0]{\catcode `\\12\catcode `\$12\catcode
  `\&12\catcode `\#12\catcode `\^12\catcode `\_12\catcode `\%12\relax}%
\providecommand \@@startlink[1]{}%
\providecommand \@@endlink[0]{}%
\providecommand \url  [0]{\begingroup\@sanitize@url \@url }%
\providecommand \@url [1]{\endgroup\@href {#1}{\urlprefix }}%
\providecommand \urlprefix  [0]{URL }%
\providecommand \Eprint [0]{\href }%
\providecommand \doibase [0]{http://dx.doi.org/}%
\providecommand \selectlanguage [0]{\@gobble}%
\providecommand \bibinfo  [0]{\@secondoftwo}%
\providecommand \bibfield  [0]{\@secondoftwo}%
\providecommand \translation [1]{[#1]}%
\providecommand \BibitemOpen [0]{}%
\providecommand \bibitemStop [0]{}%
\providecommand \bibitemNoStop [0]{.\EOS\space}%
\providecommand \EOS [0]{\spacefactor3000\relax}%
\providecommand \BibitemShut  [1]{\csname bibitem#1\endcsname}%
\let\auto@bib@innerbib\@empty
%</preamble>
\bibitem [{\citenamefont {Khintchine}(1934)}]{Khintchine1934}%
  \BibitemOpen
  \bibfield  {author} {\bibinfo {author} {\bibfnamefont {A.}~\bibnamefont
  {Khintchine}},\ }\bibfield  {title} {\enquote {\bibinfo {title}
  {{Korrelationstheorie der station{\"{a}}ren stochastischen Prozesse}},}\
  }\href {\doibase 10.1007/BF01449156} {\bibfield  {journal} {\bibinfo
  {journal} {Mathematische Annalen}\ }\textbf {\bibinfo {volume} {109}},\
  \bibinfo {pages} {604--615} (\bibinfo {year} {1934})}\BibitemShut {NoStop}%
\end{thebibliography}


\end{document}